\documentclass[fleqn,usenatbib]{mnras}

\usepackage{newtxtext,newtxmath}
\usepackage{hyperref}
\usepackage{siunitx}
\usepackage{amsmath}
\usepackage{booktabs}
\let\oldAA\AA
\renewcommand{\AA}{\text{\normalfont\oldAA}}

\usepackage[T1]{fontenc}
\usepackage{xcolor}
\DeclareRobustCommand{\VAN}[3]{#2}
\let\VANthebibliography\thebibliography
\def\thebibliography{\DeclareRobustCommand{\VAN}[3]{##3}\VANthebibliography}

\usepackage{graphicx}
\usepackage{grffile}

\title[Hierarchical black hole mergers]{Hierarchical black hole mergers in young, globular and nuclear star clusters: the effect of metallicity, spin  and cluster properties}

\author[Mapelli et al.]{Michela Mapelli,$^{1,2,3}$\thanks{E-mail: \href{michela.mapelli@unipd.it}{michela.mapelli@unipd.it}}
Marco Dall'Amico,$^{1,2}$
Yann Bouffanais,$^{1,2}$
Nicola Giacobbo,$^{1,2,3,4}$
\newauthor{Manuel Arca Sedda,$^{5}$
M. Celeste Artale,$^{6}$
Alessandro Ballone,$^{1,2,3}$
Ugo N. Di Carlo,$^{1,2}$}
\newauthor{Giuliano Iorio,$^{1,2,3}$
Filippo Santoliquido$^{1,2}$,
Stefano Torniamenti$^{1,2,3}$}
\\
$^{1}$Physics and Astronomy Department Galileo Galilei, University of Padova, Vicolo dell'Osservatorio 3, I--35122, Padova, Italy\\
$^{2}$INFN--Padova, Via Marzolo 8, I--35131 Padova, Italy\\
$^{3}$INAF--Osservatorio Astronomico di Padova, Vicolo dell'Osservatorio 5, I--35122, Padova, Italy\\
$^{4}$School of Physics and Astronomy, Institute for Gravitational Wave Astronomy, University of Birmingham, Birmingham, B15 2TT, United Kingdom\\
$^{5}$Astronomisches Rechen-Institut, Zentr\"um f\"ur Astronomie, Universit\"at Heidelberg, M\"onchofstr. 12-14, Heidelberg, Germany\\
$^{6}$Institut f{\"u}r  Astro- und Teilchenphysik, Universit{\"a}t Innsbruck, Technikerstrasse 25/8, A-6020, Innsbruck, {\"O}sterreich
}

\date{Accepted XXX. Received YYY; in original form ZZZ}

\pubyear{2020}

\begin{document}
\label{firstpage}
\pagerange{\pageref{firstpage}--\pageref{lastpage}}
\maketitle

\begin{abstract}
  
We explore hierarchical black hole (BH) mergers in nuclear star clusters (NSCs), globular clusters (GCs) and young star clusters (YSCs), accounting for both original and dynamically assembled binary BHs (BBHs). We find that the median mass of both first- and nth-generation dynamical mergers is larger in GCs and YSCs with respect to NSCs, because the lighter BHs are ejected by supernova kicks from the lower-mass clusters. Also, first- and nth-generation BH masses are strongly affected by the metallicity of the progenitor stars: the median mass of the primary BH of a nth-generation merger is $\sim{}24-38$ M$_\odot$ ($\sim{}9-15$ M$_\odot$) in metal-poor (metal-rich) NSCs. The maximum BH mass mainly depends on the escape velocity: BHs with mass up to several thousand M$_\odot$ form in NSCs, while YSCs and GCs host BHs with mass up to several hundred M$_\odot$. Furthermore, we calculate the fraction of mergers with at least one component in the pair-instability mass gap ($f_{\rm PI}$) and in the intermediate-mass BH regime ($f_{\rm IMBH}$). In the fiducial model for dynamical BBHs with metallicity $Z=0.002$, we find $f_{\rm PI}\approx{}0.05$, 0.02 and 0.007 ($f_{\rm IMBH}\approx{}0.01$, 0.002 and 0.001) in NSCs, GCs and YSCs, respectively. Both $f_{\rm PI}$ and $f_{\rm IMBH}$ drop by at least one order of magnitude at solar metallicity. Finally, we investigate the formation of GW190521 by assuming that it is either a nearly equal-mass BBH or an intermediate-mass ratio inspiral.   
\end{abstract}


\section{Introduction}

The past five years have witnessed the first three observing runs of the Advanced LIGO and Virgo gravitational-wave (GW) interferometers \citep{VIRGOdetector,LIGOdetector}, leading to the detection of about 50 binary compact object mergers \citep{abbottGW150914,abbottastrophysics,abbottO1,abbottGW170817,abbottO2,abbottO2popandrate,abbottO3a,abbottO3popandrate,abbottGW190425,abbottGW190412,abbottGW190814}. Based on the results of independent
pipelines, \cite{zackay2019}, \cite{venumadhav2020}  and \cite{nitz2020}
claimed several additional GW candidates. 
This growing sample represents a Rosetta stone to investigate the formation of binary compact objects.

Among the main results of the third observing run, GW190521 \citep{abbottGW190521,abbottGW190521astro} is the most massive binary black hole (BBH) merger observed to date. The primary component mass of GW190521 ($\approx{}85$ M$_\odot$) challenges current models of stellar evolution and pair instability (PI) supernovae \citep[e.g.,][]{belczynski2016pair,woosley2017,woosley2019,spera2017,farmer2019,farmer2020,mapelli2020,tanikawa2020,farrell2020,belczynski190521,costa2020}. \cite{nitz2021} proposed a different interpretation for GW190521, as an intermediate-mass ratio inspiral involving an intermediate-mass black hole (IMBH) and a $\sim{}16$ M$_\odot$ companion (see also \citealt{fishbach2020}).

Several channels can lead to the formation of BBHs: i) pairing of primordial black holes \citep[e.g.,][]{carr1974,carr2016,bird2016,ali2017,scelfo2018,deluca2021}, ii) binary star evolution through common envelope \citep[e.g.,][]{tutukov1973, bethe1998,portegieszwart1998,belczynski2002, belczynski2008,dominik2013, belczynski2016,eldridge2016,stevenson2017,mapelli2017,mapelli2018,mapelli2019, klencki2018,kruckow2018,spera2019,neijssel2019,eldridge2019,tang2019} or via homogeneous mixing \citep[e.g.,][]{marchant2016,demink2016,mandel2016,dubuisson2020,riley2020}, and iii) dynamical processes in triples \citep[e.g.,][]{antonini2016b,antonini2017,arcasedda2018b,fragione2019,fragione2020,vigna2021}, young star clusters (YSCs, e.g., \citealt{banerjee2010,ziosi2014,mapelli2016,banerjee2017,banerjee2018,banerjee2020,dicarlo2019,dicarlo2020b,kumamoto2019,kumamoto2020}), globular clusters (GCs, e.g., \citealt{portegieszwart2000,downing2010,rodriguez2015,rodriguez2016,rodriguez2018,askar2017,samsing2014,samsing2018,fragione2018,zevin2019,roupas2019,antonini2020}), nuclear star clusters (NSCs, e.g., \citealt{oleary2009,miller2009,antonini2016,petrovich2017,rasskazov2019,arcasedda2018,arcasedda2019,arcasedda2020,arcasedda2020b}) and AGN discs \citep[e.g.,][]{mckernan2012,mckernan2018,bartos2017,stone2017,yang2019,tagawa2020}.

One of the distinctive signatures of the dynamical scenario is the formation of hierarchical mergers, i.e. repeated mergers of stellar-origin black holes (BHs) that build up more massive ones \citep{miller2002,fishbach2017,gerosa2017,doctor2020,belczynskibanerjee2020,flitter2020}. This process is possible only in dense star clusters, where the merger remnant, which is initially a single BH, can pair up by dynamical exchanges or three-body encounters \citep[e.g.,][]{heggie1975,hills1980}.   The main obstacle to the formation of second-generation BHs via hierarchical mergers is the relativistic kick that the merger remnant receives at birth 
(e.g., \citealt{fitchett1983,favata2004,campanelli2007,lousto2011}). This kick can be up to several thousand km s$^{-1}$ and can easily eject the BH remnant from its parent star cluster \citep{holley-bockelmann2008,moody2009,fragione2018b,gerosa2019,arcasedda2020}. Hence, the interplay between the properties of the host star cluster (e.g., its escape velocity), those of the first-generation BBH population and the magnitude of the kick decides the maximum mass of a merger remnant in a given environment \citep[e.g.,][]{rodriguez2019,kimball2020}.

Due to their high escape velocity ($v_{\rm esc}\sim{}100$~km~s$^{-1}$), NSCs are more likely to retain hierarchical mergers than other star clusters (e.g., \citealt{antonini2016,yang2019,arcasedda2019c,arcasedda2020}). \cite{antonini2019} find that BH growth becomes substantial for $v_{\rm esc}>300$~km~s$^{-1}$, leading to the formation of  IMBHs (see also \citealt{fragione2020}).
Hence, hierarchical mergers can build up IMBHs and also partially fill the PI mass gap between $\sim{}65$ and $\sim{}120$ M$_\odot$, explaining the formation of BBHs like GW190521 \citep[e.g.,][]{fragione2020b}.

One of the main challenges of studying hierarchical mergers is the computational cost. It is nearly impossible to investigate the relevant parameter space with hybrid Monte Carlo and/or N-body simulations of massive star clusters, especially GCs and NSCs. Here, we present a new fast and flexible semi-analytic model to investigate hierarchical mergers in different environments, complementary to dynamical simulations. Our new tool allows us to probe the parameter space, including BBH masses, spins, delay times, orbital eccentricities, and star cluster properties. With respect to previous semi-analytic work \citep[e.g.,][]{choksi2019,arcasedda2020,baibhav2020,fragione2020}, our new model fully integrates the evolution of the orbital properties (semi-major axis and eccentricity) by taking into account both GW emission and three-body encounters. Moreover, we consider both original BBHs, i.e. BBHs originating from binary stars, and dynamical BBHs, i.e. BBHs formed by dynamical pairing.  Our tool consists in a suite of python scripts, dubbed {\sc fastcluster}, and is available upon request. 

This paper is organized as follows. In Section~\ref{sec:firstgen}, we lay down the properties of first generation BBHs.  Sections~\ref{sec:evolution} and \ref{sec:nthgen} describe the orbital evolution and the properties of nth-generation BBHs, respectively. Section~\ref{sec:starcluster} outlines the relevant star cluster properties. Sections~\ref{sec:results} and \ref{sec:discussion} describe and discuss our main results. Finally, Section~\ref{sec:summary} is a summary of the paper.

\section{First-generation BBHs}\label{sec:firstgen} 

In the following analysis, we assume that BBH mergers in star clusters belong to two channels: i) original binaries, i.e. BBHs that originate from binary stars, and ii) dynamical binaries, i.e. BBHs that form from three-body encounters. The contribution of original binaries has often been neglected in the past, because, in the densest and most massive star clusters,  a relevant fraction of the original binaries are ionized before they can form a BBH  \citep[e.g.,][]{morscher2015,antonini2016}. However, especially in low-mass clusters ($\leq{}10^4$ M$_\odot$), original binaries can survive and contribute to $\ge{}10$~\% of the simulated BBH mergers \citep[e.g.,][]{dicarlo2019}.

\begin{table}
	\begin{center}
	\caption{Summary of the models.}
	\label{tab:table1}
	\begin{tabular}{cccccc} 
		\toprule
                Model & Formation & Star Cluster & $Z$  & $\sigma_{\chi}$ & $\sigma_{\rm M}$\\		
		\midrule
 NSC\_D1 & Dynamical & NSC & 0.002 & 0.1 & 0.4\\
 NSC\_D2 & Dynamical & NSC & 0.0002 & 0.1 & 0.4\\
 NSC\_D3 & Dynamical & NSC & 0.02 & 0.1 & 0.4\\
 NSC\_D4 & Dynamical & NSC & 0.002 & 0.01 & 0.4\\
 NSC\_D5 & Dynamical & NSC & 0.002 & 0.3 & 0.4\\
 NSC\_D6 & Dynamical & NSC & 0.002 & 0.1 & 0.2\\
 NSC\_D7 & Dynamical & NSC & 0.002 & 0.1 & 0.6\\
 GC\_D1 & Dynamical & GC & 0.002 & 0.1 & 0.4\\
 GC\_D2 & Dynamical & GC & 0.0002 & 0.1 & 0.4\\
 GC\_D3 & Dynamical & GC & 0.02 & 0.1 & 0.4\\
 GC\_D4 & Dynamical & GC & 0.002 & 0.01 & 0.4\\
 GC\_D5 & Dynamical & GC & 0.002 & 0.3 & 0.4\\
 GC\_D6  & Dynamical & GC & 0.002 & 0.1 & 0.2\\
 GC\_D7 & Dynamical & GC & 0.002 & 0.1 & 0.6\\
 YSC\_D1 & Dynamical & YSC & 0.002 & 0.1 & 0.4\\
 YSC\_D2 & Dynamical & YSC & 0.0002 & 0.1 & 0.4\\
 YSC\_D3 & Dynamical & YSC & 0.02 & 0.1 & 0.4\\
 YSC\_D4 & Dynamical & YSC & 0.002 & 0.01 & 0.4\\
 YSC\_D5  & Dynamical & YSC & 0.002 & 0.3 & 0.4\\
 YSC\_D6 & Dynamical & YSC & 0.002 & 0.1 & 0.2\\
 YSC\_D7 & Dynamical & YSC & 0.002 & 0.1 & 0.6\vspace{0.1cm}\\
NSC\_O1 & Original & NSC & 0.002 & 0.1 & 0.4\\
 NSC\_O2 & Original & NSC & 0.0002 & 0.1 & 0.4\\
 NSC\_O3 & Original & NSC & 0.02 & 0.1 & 0.4\\
 NSC\_O4 & Original & NSC & 0.002 & 0.01 & 0.4\\
 NSC\_O5 & Original & NSC & 0.002 & 0.3 & 0.4\\
 NSC\_O6 & Original & NSC & 0.002 & 0.1 & 0.2\\
 NSC\_O7 & Original & NSC & 0.002 & 0.1 & 0.6\\
 GC\_O1 & Original & GC & 0.002 & 0.1 & 0.4\\
 GC\_O2 & Original & GC & 0.0002 & 0.1 & 0.4\\
 GC\_O3 & Original & GC & 0.02 & 0.1 & 0.4\\
 GC\_O4 & Original & GC & 0.002 & 0.01 & 0.4\\
 GC\_O5 & Original & GC & 0.002 & 0.3 & 0.4\\
 GC\_O6  & Original & GC & 0.002 & 0.1 & 0.2\\
 GC\_O7 & Original & GC & 0.002 & 0.1 & 0.6\\
 YSC\_O1 & Original & YSC & 0.002 & 0.1 & 0.4\\
 YSC\_O2 & Original & YSC & 0.0002 & 0.1 & 0.4\\
 YSC\_O3 & Original & YSC & 0.02 & 0.1 & 0.4\\
 YSC\_O4 & Original & YSC & 0.002 & 0.01 & 0.4\\
 YSC\_O5  & Original & YSC & 0.002 & 0.3 & 0.4\\
 YSC\_O6 & Original & YSC & 0.002 & 0.1 & 0.2\\
 YSC\_O7 & Original & YSC & 0.002 & 0.1 & 0.6 \vspace{0.1cm}\\
 \bottomrule
	\end{tabular}
	\end{center}
	\footnotesize{Column 1: Name of the model (NSC\_D1, GC\_D1 and YSC\_D1 are our fiducial models for dynamical binaries, while NSC\_O1, GC\_O1 and YSC\_O1 are our fiducial models for original binaries); column 2: Formation path of the binary (original or dynamical); column 3: Star cluster type (NSC, GC, YSC); column 4: metallicity of first-generation BHs ($Z=0.0002$, 0.002, 0.02); column 5: root-mean square value of the Maxwellian distribution of spin magnitudes ($\sigma_\chi{}=$ 0.01, 0.1, 0.3); column 6: standard deviation of the log-normal distribution of total star cluster masses ($\sigma_{\rm M}=0.2,$ 0.4, 0.6).}
\end{table}

\subsection{Original binaries}\label{sec:originalbin}

We generate catalogues of mass, semi-major axis and eccentricity of original BBHs with our population-synthesis code\footnote{{\sc mobse} is publicly available at \url{http://demoblack.com/}.} {\sc mobse} \citep{mapelli2017,giacobbo2018}. 
These catalogues represent the initial properties of our first-generation population. However, the {\sc fastcluster} code is extremely flexible and can take its input catalogues from different models, including phenomenological ones.

To build the {\sc mobse} catalogues, we use the same set-up as run $\alpha{}5$MT0.5 in \cite{santoliquido2020b}. In particular, we assume a common envelope parameter $\alpha{}=5$ and a mass transfer efficiency $f_{\rm MT}=0.5$. 
We adopt the delayed model by \cite{fryer2012} for core-collapse supernovae, while for (pulsational) PI supernovae we use the equations reported in the appendix of \cite{mapelli2020}. This yields a minimum BH mass of $\approx{}3$ M$_\odot$ and a maximum BH mass of $\approx{}65$ M$_\odot$ from single star evolution \citep{giacobbo2018}. In tight binary systems, the maximum BH mass is lower ($\approx{}45$ M$_\odot$), because the common envelope process removes any residual hydrogen envelope, reducing the final BH mass \citep{giacobbo2018b}. 

We draw the zero-age main sequence mass of the primary stars from a Kroupa \citep{kroupa2001} initial mass function. The initial orbital parameters (semi-major axis, orbital eccentricity and mass ratio) of binary stars have been randomly drawn as already described in \cite{santoliquido2020b}. In particular, we derive the mass ratio $q=m_2/m_1$ as $\mathcal{F}(q) \propto q^{-0.1}$ with $q\in [0.1,\,{}1]$, the orbital period $P$ from $\mathcal{F}(\Pi) \propto \Pi^{-0.55}$ with $\Pi = \log{(P/\text{day})} \in [0.15,\,{} 5.5]$ and the eccentricity $e$ from $\mathcal{F}(e) \propto e^{-0.42}~~\text{with}~~ 0\leq e \leq 0.9$ \citep{sana2012}.   To generate our BBH catalogues, we ran $2\times{}10^7$, $10^7$ and $10^7$ binary stars  at metallicity $Z=0.02$, 0.002 and 0.0002, respectively. We simulated more binary systems at $Z=0.02$ because of the lower merger efficiency \citep[e.g.,][]{giacobbo2018b,klencki2018}. The case with $Z=0.02$ is approximately the solar metallicity \citep{anders1989}.

Our original BBHs are uniformly sampled from the BBHs that merge within a Hubble time in the {\sc mobse} catalogues. From the {\sc mobse} catalogues, we take the masses of the two BHs, the value of the semi-major axis and that of the orbital eccentricity at the time of the formation of the second BH ($t_{\rm form}$). We assume that the binary star evolved nearly unperturbed before $t_{\rm form}$. This is justified by the extremely short orbital period of these binaries (mostly $\leq{}50$ R$_\odot$). In a follow-up study, we will integrate the dynamical evolution of the progenitor binary stars inside {\sc fastcluster}.

We draw the dimensionless spin magnitudes of the primary and secondary BHs ($\chi_1$ and $\chi_2$) from a Maxwellian distribution with one-dimensional root-mean square $\sigma{}_\chi=0.01,\,{}0.1$ and 0.3 and truncated at $\chi{}=1$. $\sigma{}_\chi=0.1$ is our fiducial case, because it is quite reminiscent of the spins inferred from the second GW transient catalogue \citep[GWTC-2, see Figure 10 of ][]{abbottO3popandrate}. We draw the direction of the spins isotropic over the sphere. 
We make this assumption because even the original BBHs interact with other stars in a star cluster: even if they do not undergo dynamical exchanges, close encounters should heavily affect their spin orientations \cite[][]{trani2021,kumamoto2021}. For each binary, we also estimate the effective spin ($\chi_{\rm eff}$) and the precessing spin ($\chi_{\rm p}$), which are defined as follows:
\begin{eqnarray}
\chi_{\rm eff}= \frac{(m_1\,{}\vec{\chi}_1+m_2\,{}\vec{\chi}_2)}{m_1+m_2}\cdot{}\frac{\vec{L}}{L},
\nonumber{}\\
\chi_{\rm p}=\frac{c}{B_1\,{}G\,{}m_1^2}\,{}\max{(B_1\,{}S_{1\perp{}},\,{}B_2\,{}S_{2\,{}\perp})},
\end{eqnarray}
where $m_1$ ($m_2$) is the primary (secondary) BH mass, $\vec{L}$ is the orbital angular momentum of the BBH, $c$ is the speed of light, $G$ is the gravity constant, $B_1\equiv{}2+3\,{}q/2$ and $B_2\equiv{}2+3/(2\,{}q)$, with $q=m_2/m_1$.  $S_{1\perp{}}$ and $S_{2\perp{}}$ are the components of the spin vectors ($\vec{S}_1$ and $\vec{S}_2$) perpendicular to the orbital angular momentum. Finally $\vec{\chi}_1\equiv{}c\,{}\vec{S}_1/(G\,{}m_1^2)$ and $\vec{\chi}_2\equiv{}c\,{}\vec{S}_2/(G\,{}m_2^2)$.

An original binary survives in a star cluster only if it is hard, i.e. if its binding energy $E_{\rm b}$ is larger than the average kinetic energy of a star $\langle{}E_{\rm k}\rangle{}$ \citep{heggie1975}. Hence, out of all the BBHs calculated with {\sc mobse}, we dynamically evolve only those for which 
   \begin{equation}\label{eq:hb}
    E_{\rm b}=\frac{G\,{}m_1\,{}m_2}{2\,{}a}\ge{}\langle{}E_{\rm k}\rangle{}=\frac{1}{2}\,{}m_\ast\,{}\sigma{}^2,
  \end{equation}
  where $a$ is the semi-major axis,    $m_\ast$ is the average mass of a star in the star cluster and $\sigma{}$ is the three-dimensional velocity dispersion.

  If an original BBH does not satisfy the condition in eq.~\ref{eq:hb}, we do not consider it any further. Otherwise, we evolve it by hardening and by GW emission, as described in Section~\ref{sec:evolution}.

\subsection{Dynamical binaries}\label{sec:dynbin}

Dynamical binaries form either via exchanges between a binary star and an intruder, or via dynamical encounters of three initially single bodies (e.g., \citealt{heggie1975,hills1980}). The latter are favored when the fraction of binary systems is small and the local density is extremely high (e.g., during a core collapse), while the former are dominant in star clusters with a large binary fraction, such as YSCs.

The first important requirement for a single BH to pair up with another BH is that it has reached the dense core of the star cluster, where three-body encounters are more likely. This happens over a dynamical friction timescale \citep{chandrasekhar1943}:
\begin{equation}\label{eq:tdf}
  t_{\rm DF}=\frac{3}{4\left(2\,{}\pi{}\right)^{1/2}\,{}G^2\ln{\Lambda{}}}\,{}\frac{\sigma^3}{m_{\rm BH}\,{}\rho{}},
\end{equation}
where $m_{\rm BH}$ is the mass of the BH, $\sigma{}$ is the 3D velocity dispersion, $\rho{}$ is the mass density at the half-mass radius and $\ln{}\Lambda{}\sim{}10$ is the Coulomb logarithm. After a time $t_{\rm DF}$, the BH has sunk to the core of the cluster and can acquire a companion by three-body encounters or by exchange.

The timescale for binary formation via the encounter of three single BHs is \citep{goodman1993,lee1995,fragione2020}:
\begin{eqnarray}\label{eq:t3bb}
  t_{\rm 3bb}=125\,{}{\rm Myr}\,{}\left(\frac{10^6\,{}{\rm pc}^{-3}}{n_{\rm c}}\right)^2\,{}\left(\zeta{}^{-1}\,{}\frac{\sigma_{\rm 1D}}{30\,{}{\rm km}\,{}{\rm s}^{-1}}\right)^9\,{}\left(\frac{20\,{}{\rm M}_\odot}{m_{\rm BH}}\right)^5,
\end{eqnarray}
where  $n_{\rm c}$ is the central density of the star cluster, $\sigma{}_{\rm 1D}=\sigma{}/\sqrt{3}$ is the one-dimensional velocity dispersion at the half-mass radius (assuming isotropic distribution of stellar velocities) and
\begin{equation}\label{eq:zeta}
\zeta{}=\frac{m_\ast{}\,{}\sigma{}^2}{m_{\rm BH}\,{}\sigma_{\rm BH}^2}.
\end{equation}
In eq.~\ref{eq:zeta}, $\sigma_{\rm BH}$ is the velocity dispersion associated with massive BHs of mass $m_{\rm BH}$. In case of equipartition, $\zeta{}=1$. If the system is not in equipartition (i.e., Spitzer instability takes place, \citealt{spitzer1969}), then $\zeta{}<1$. Here, we assume $\zeta{}=1$. 
 The strong dependence of eq.~\ref{eq:t3bb} on $\sigma_{\rm 1D}$ and $m_{\rm BH}$ makes it critical for the formation of BBHs in dense stellar systems. Here, we use the formalism discussed in \cite{oleary2006} and \cite{morscher2015}, which has been adopted in hybrid Monte Carlo simulations  and compares well with direct N-body simulations of globular clusters \citep{rodriguez2016compare}.

Finally, the timescale for the dynamical exchange of a BH into a binary star is \citep{millerlauburg2009}
\begin{eqnarray}\label{eq:t12}
  t_{\rm 12}=3\,{}{\rm Gyr}\,{}\left(\frac{0.01}{f_{\rm bin}}\right)\,{}\left(\frac{10^6\,{}{\rm pc}^{-3}}{n_{\rm c}}\right)\,{}\left(\frac{\sigma}{50\,{}{\rm km}\,{}{\rm s}^{-1}}\right) \nonumber\\
  \,{}\left(\frac{12\,{}{\rm M}_\odot}{m_{\rm BH}+2\,{}m_\ast}\right)\,{}\left(\frac{1\,{}{\rm AU}}{a_{\rm hard}}\right),
\end{eqnarray}
where $f_{\rm bin}$ is the binary fraction and $a_{\rm hard}=G\,{}m_\ast/\sigma{}^2$ is the minimum semi-major axis of a hard binary system.

We assume that dynamical BBHs form at a time 
\begin{equation}\label{eq:tdyn}
  t_{\rm dyn}=\max{\left\{t_{\rm form},\left[t_{\rm DF}+\min{(t_{\rm 3bb},t_{\rm 12})}\right]\right\}},
\end{equation}
where $t_{\rm form}$ is 
the time for the collapse of the progenitor star to a BH, while $t_{\rm DF}$, $t_{\rm 3bb}$ and $t_{\rm 12}$ are the timescales defined by eqs.~\ref{eq:tdf}, \ref{eq:t3bb} and \ref{eq:t12}, respectively. In equation~\ref{eq:tdyn}, we assume that not only the BH but also its progenitor star undergoes dynamical friction, exchanges and three-body encounters.

In the densest clusters, we should also consider the timescale $t_{\rm cap}$ for GW two-body captures (e.g., \citealt{quinlan1990}):
\begin{eqnarray}
  t_{\rm cap}\sim{}7.7\times{}10^{3}\,{}{\rm Gyr}\,{}\left(\frac{{\rm M}_\odot}{m_{\rm BH}}\right)^{2} 
  \,{}\left(\frac{10^8\,{}{\rm pc}^{-3}}{n_{\rm c}}\right)\left(\frac{\sigma{}}{200\,{}{\rm km\,{}s}^{-1}}\right)^{11/7}.
  \nonumber\\
\end{eqnarray}
However, this timescale is always longer than $t_{\rm 3bb}$ and $t_{\rm 12}$ for all the simulated clusters. Moreover, we also neglect the timescale for binary--binary encounters which might be even shorter than $t_{\rm 12}$ in star clusters with a high binary fraction \citep[e.g.,][]{zevin2019}.

We randomly draw the mass $m_1$ of the primary BH of first-generation dynamical BBHs from the list of single BHs and BHs in loose binaries in the same {\sc mobse} catalogues as we have described in Section~\ref{sec:originalbin}. BHs from single stars and loose binaries come from the same mass distribution function, because only mass transfer and common envelope can affect the final masses in a binary system. 
The mass $m_2$ of the secondary BH is randomly drawn in the interval $[m_{\rm min},\,{}m_1)$, where $m_{\rm min}=3$ M$_\odot$, following the probability distribution function \citep{oleary2016}:
  \begin{equation}\label{eq:oleary}
    p(m_2)\propto{}(m_1+m_2)^4.
\end{equation}

  After generating each mass, we check that both the primary and the secondary BH are not ejected from the star cluster by the supernova kick.  We calculate the kick as
  \begin{equation}
    v_{\rm SN}=v_{\rm H05}\,{}\frac{\langle{}m_{\rm NS}\rangle{}}{m_{\rm BH}},
  \end{equation}
  where $\langle{}m_{\rm NS}\rangle{}=1.33$ M$_\odot$  is the average neutron star mass \citep{oezel2016}, $m_{\rm BH}$ is the mass of the BH, and $v_{\rm H05}$ is a number randomly drawn from a Maxwellian distribution with one-dimensional root-mean square $\sigma_V=265$ km s$^{-1}$. This distribution matches the proper motions of young Galactic pulsars according to the fit by \cite{hobbs2005} and is adapted to BHs based on linear momentum conservation.   If the primary or the secondary BH have $v_{\rm SN}>v_{\rm esc}$ (where $v_{\rm esc}$ is the escape velocity from the star cluster), we reject that binary and we draw a new one as detailed above.

  We generate the spin magnitude and direction of dynamical BBHs in the same way as those of original BBHs (Section~\ref{sec:originalbin}). We draw the initial eccentricities of dynamical BBHs from the thermal probability distribution $p(e)=2\,{}e$ with $e\in{}[0,\,{}1)$ \citep{heggie1975}. The initial semi-major axes follow a distribution $p(a)\propto{}1/a$ with $a_{\rm min}=1$ R$_\odot$ and $a_{\rm max}=10^3$ R$_\odot$. If the binary generated in this way is soft according to eq.~\ref{eq:hb}, we reject the value of $a$ and we generate a new value. In this way, all the dynamical binaries are initially hard binaries. We then evolve each dynamical binary as described in Section~\ref{sec:evolution}.

\begin{figure*}
  \begin{center}
    \includegraphics[width = 0.85 \textwidth]{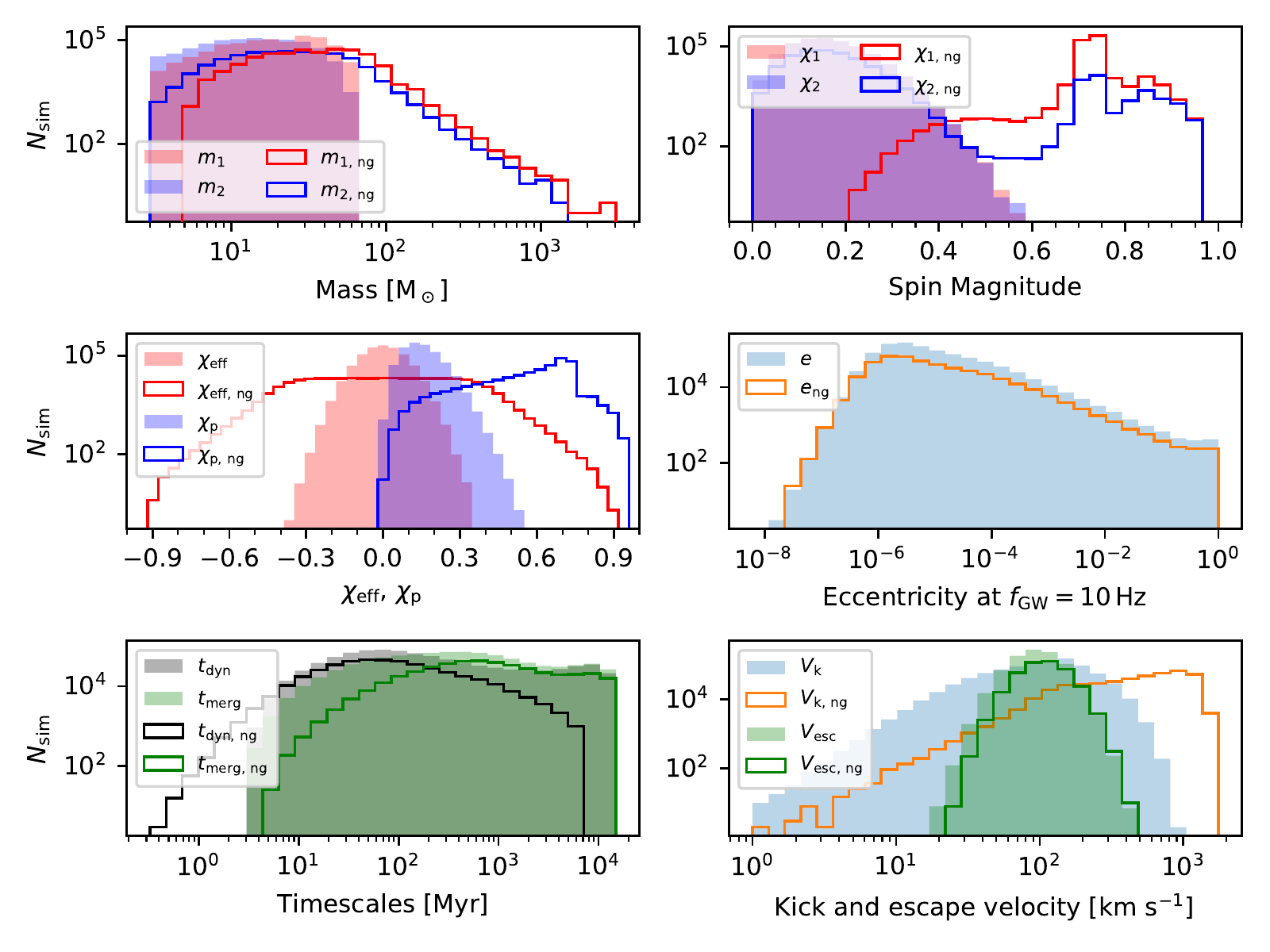}
    \end{center}
  \caption{Main properties of dynamical BBH mergers in NSCs, according to our fiducial model (NSC\_D1). The {\bf upper left-hand panel} shows the relevant masses. Filled red (blue) histogram: mass of the primary, $m_1$ (secondary,  $m_2$), in first-generation mergers; unfilled red (blue) histogram: mass of the primary, $m_{\rm 1,\,{}ng}$ (secondary, $m_{\rm 2,\,{}ng}$) in nth-generation mergers. The {\bf upper right-hand panel} shows the dimensionless spin parameters. Filled red (blue) histogram: dimensionless spin of the primary, $\chi_1$  (secondary, $\chi_2$) in first-generation mergers; unfilled red (blue) histogram: dimensionless spin of the primary, $\chi_{\rm 1,\,{}ng}$ (secondary, $\chi_{\rm 2,\,{}ng}$) in nth-generation mergers. The {\bf middle left-hand panel} shows the effective and precessing spin parameters. Filled red (blue) histogram: effective (precessing) spin $\chi_{\rm eff}$ ($\chi_{\rm p}$) in first-generation mergers; unfilled red (blue) histogram: effective (precessing) spin $\chi_{\rm eff,\,{}ng}$ ($\chi_{\rm p,\,{}ng}$) in nth-generation mergers. In the {\bf middle right-hand panel}, filled light-blue (unfilled orange) histogram: orbital eccentricity when the GW frequency $f_{\rm GW}=10$ Hz for first-generation (nth-generation) mergers. In the {\bf lower left-hand panel}, we show the most important timescales. Gray filled histogram: time for the dynamical formation of first-generation BBHs ($t_{\rm dyn}$, eq.~\ref{eq:tdyn}); green filled histogram: delay time for the merger of the fist-generation BBHs ($t_{\rm merg}$); black unfilled histogram:  time for the dynamical formation of nth-generation BBHs ($t_{\rm dyn,\,{}ng}$, eq.~\ref{eq:tdynng}); green unfilled histogram: delay time for the merger of the nth-generation BBHs ($t_{\rm merg,\,{}ng}$). The {\bf lower right-hand panel} shows the most important velocities. Light-blue filled histogram: relativistic kick velocity received by the merger product of the first-generation BBHs ($V_{\rm k}$); orange unfilled histogram: relativistic kick velocity received by the merger product of the nth-generation BBHs ($V_{\rm k,\,{}ng}$); green filled histogram: escape velocity of the star clusters that host the first-generation BBH mergers ($V_{\rm esc}$); green unfilled histogram: escape velocity of the star clusters that host the nth-generation BBH mergers ($V_{\rm esc,\,{}ng}$). The $y-$axis of all the histograms shows the number of simulated BHs $N_{\rm sim}$, without normalization. 
    \label{fig:NSCdyn}}
\end{figure*}

\begin{figure*}
  \begin{center}
    \includegraphics[width = 0.85 \textwidth]{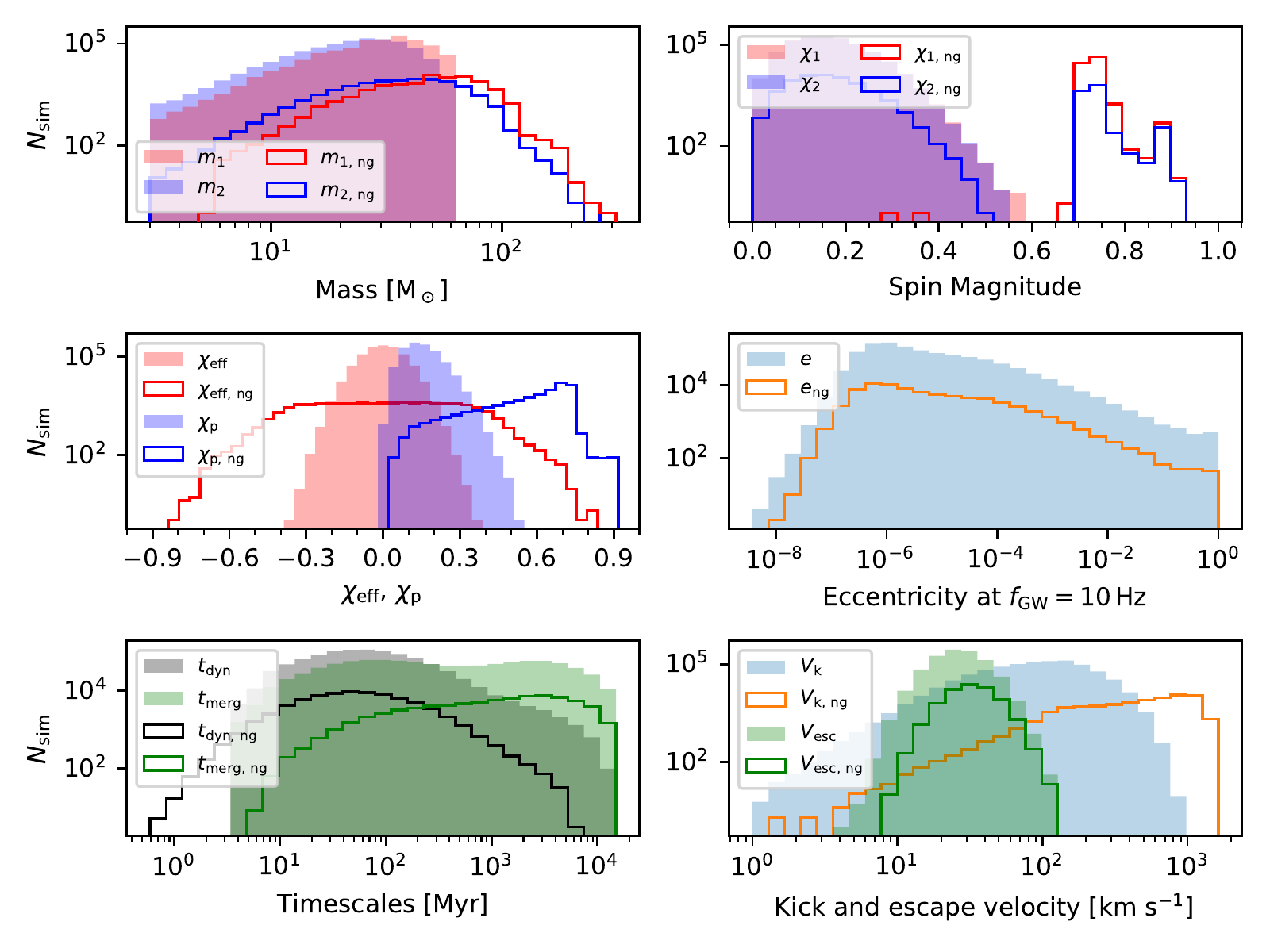}
    \end{center}
  \caption{Same as Figure~\ref{fig:NSCdyn}, but for GCs, according to our fiducial model (GC\_D1). 
    \label{fig:GCdyn}}
\end{figure*}

\begin{figure*}
  \begin{center}
    \includegraphics[width = 0.85 \textwidth]{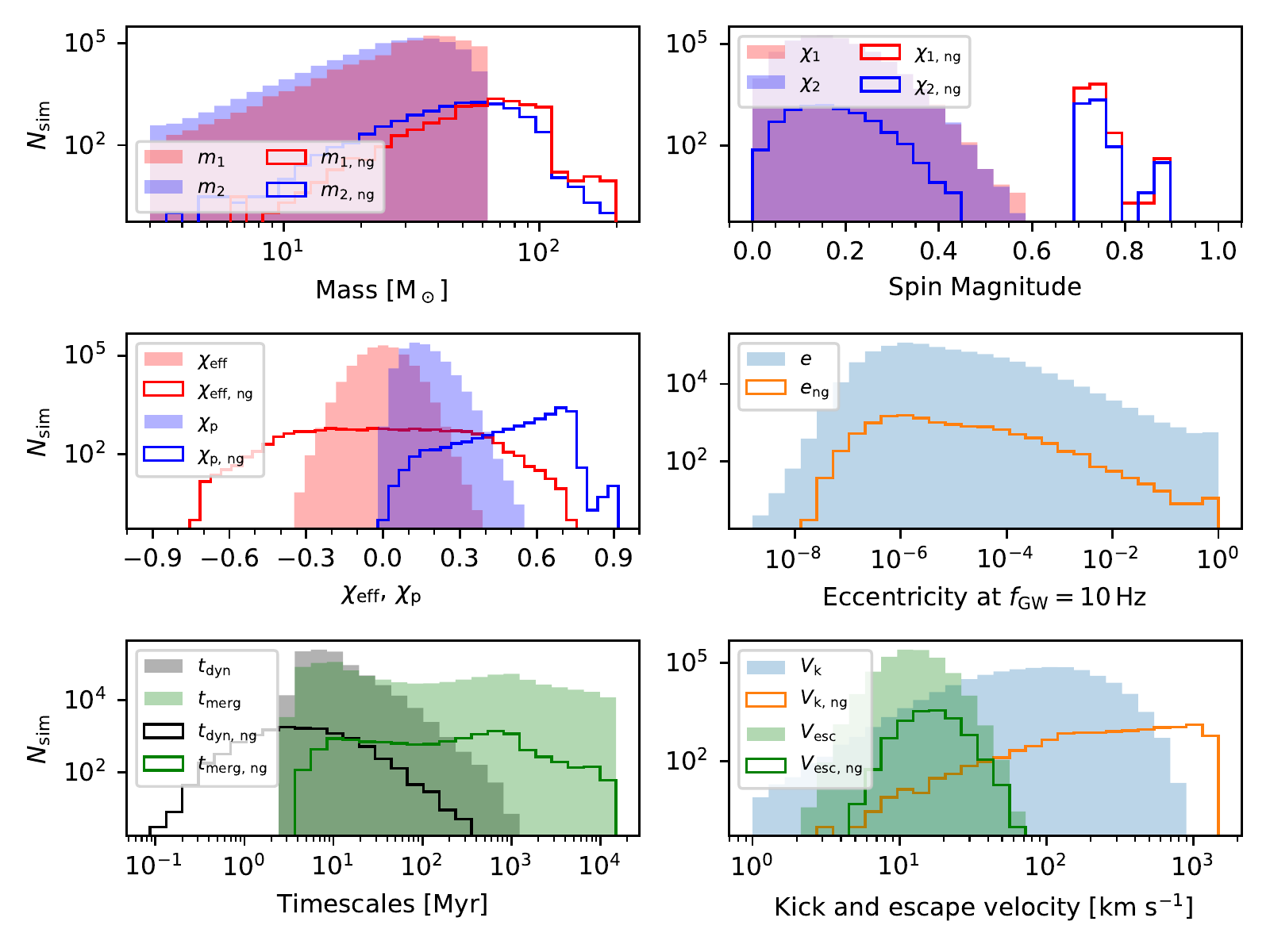}
    \end{center}
  \caption{Same as Figure~\ref{fig:NSCdyn}, but for YSCs, according to our fiducial model (YSC\_D1). 
    \label{fig:YSCdyn}}
\end{figure*}

\begin{figure*}
  \begin{center}
    \includegraphics[width = 0.85 \textwidth]{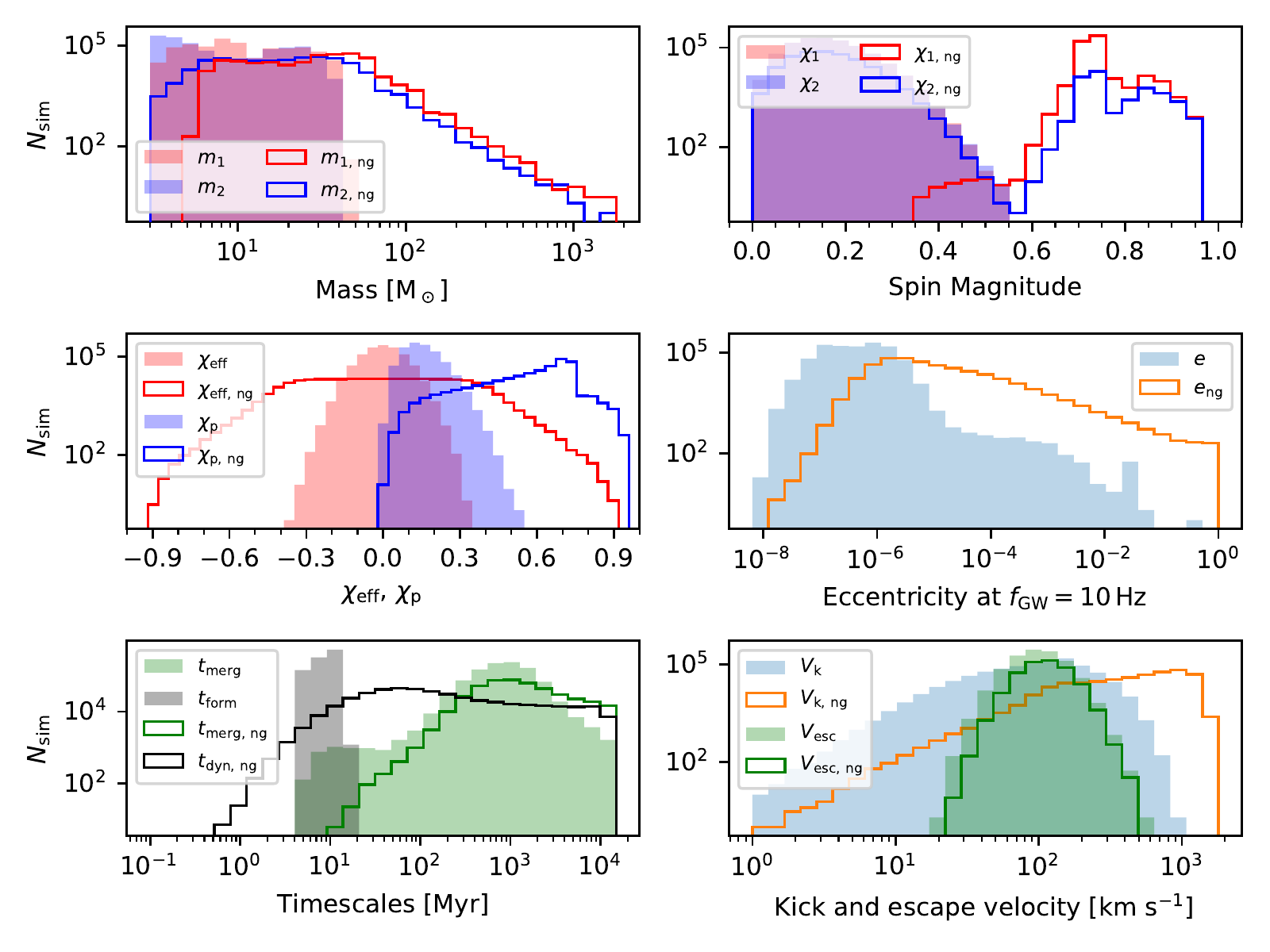}
    \end{center}
  \caption{Same as Figure~\ref{fig:NSCdyn}, but for original binaries in NSCs (model NSC\_O1). The only difference is the meaning of the filled gray histogram in the lower left-hand panel, which here shows the formation timescale of the original first-generation BBH ($t_{\rm form}$), i.e. the time elapsed between the formation of the binary star and the formation of the second BH.
    \label{fig:NSCoriginal}}
\end{figure*}

\begin{figure*}
  \begin{center}
    \includegraphics[width = 0.85 \textwidth]{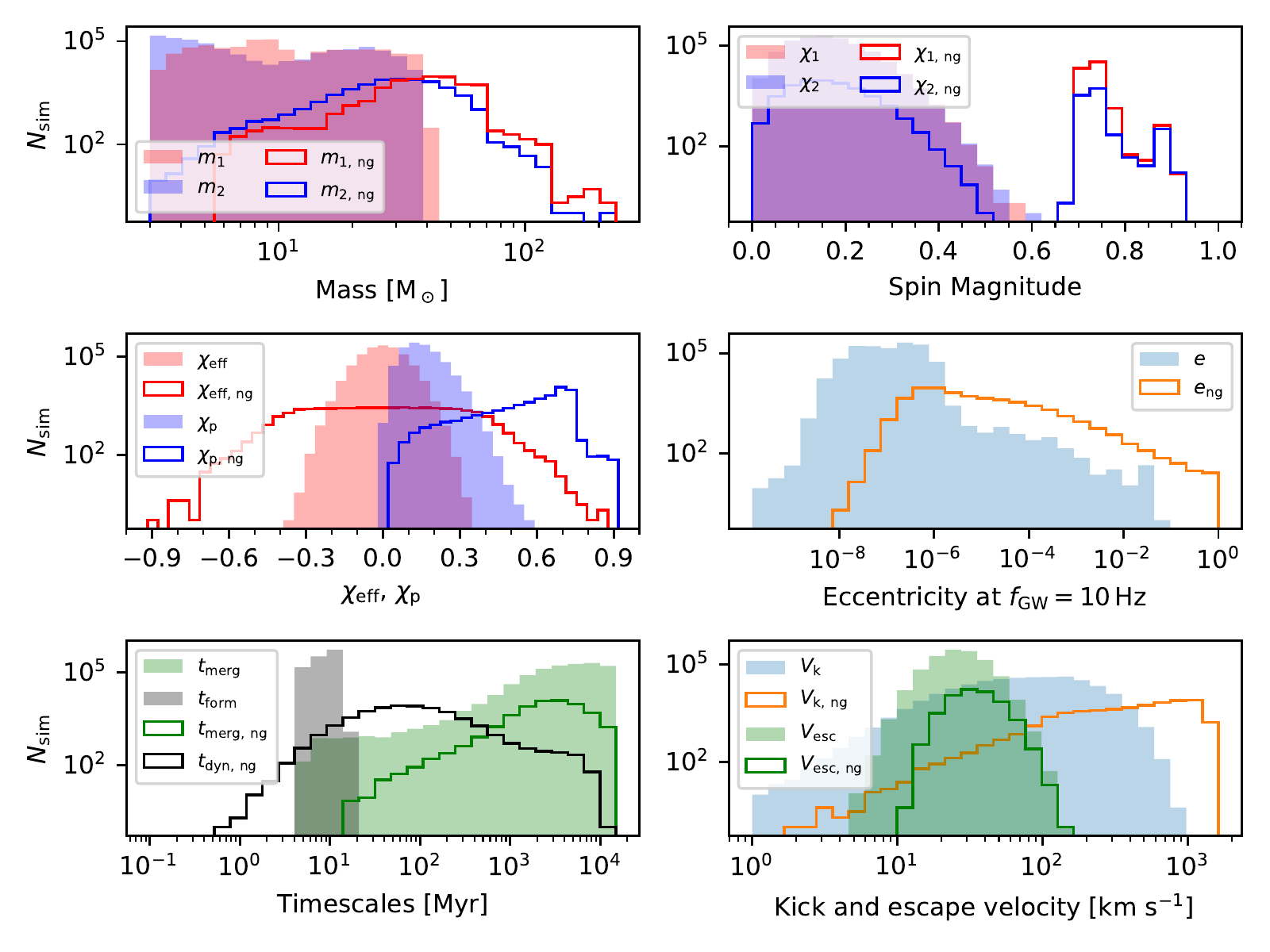}
    \end{center}
  \caption{Same as Figure~\ref{fig:NSCoriginal}, but for original binaries in GCs. Model GC\_O1. 
    \label{fig:GCoriginal}}
\end{figure*}

\begin{figure*}
  \begin{center}
    \includegraphics[width = 0.85 \textwidth]{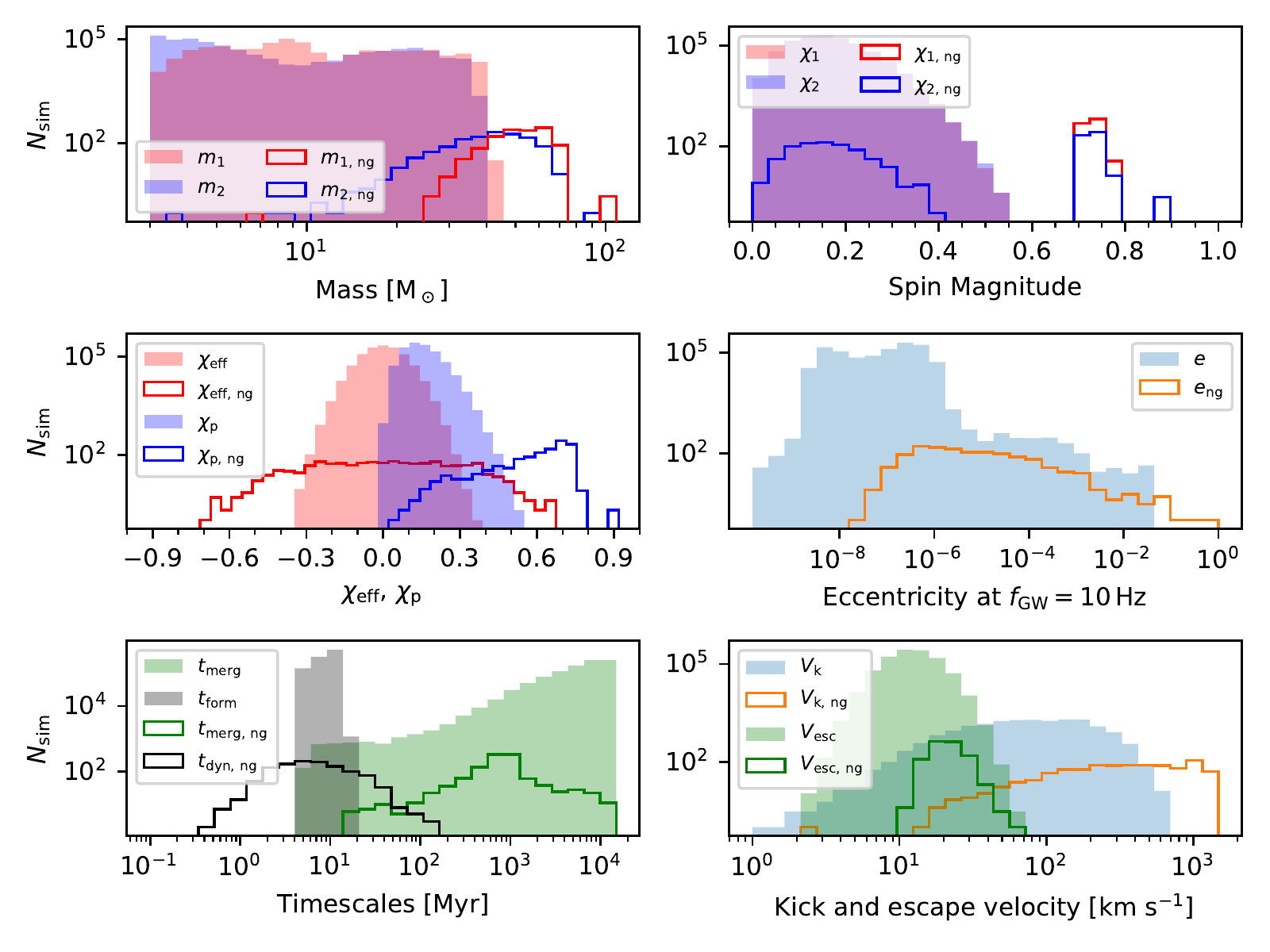}
    \end{center}
  \caption{Same as Figure~\ref{fig:NSCoriginal}, but for original binaries in YSCs. Model YSC\_O1.
    \label{fig:YSCoriginal}}
\end{figure*}
    
\section{Orbital evolution}\label{sec:evolution}

After their formation, both original  and dynamical BBHs undergo binary-single encounters, i.e. close Newtonian gravitational encounters with single stars or BHs in the star cluster. A hard binary hardens by binary-single encounters, i.e. its semi-major axis decreases at a rate \citep{heggie1975}:
\begin{equation}\label{eq:hardening}
\frac{{\rm d}a}{{\rm d}t}=-2\,{}\pi{}\,{}\xi{}\,{}\frac{G\,{}\rho{}_{\rm c}}{\sigma}\,{}a^2,
\end{equation}
where $\rho{}_{\rm c}$ is the core mass density and $\xi{}\sim{}0.1-10$ is a numerically calibrated constant \citep{hills1983}. \cite{quinlan1996} find $\xi{}\approx{}3$, while \cite{miller2002} and \cite{sesana2006} obtain $\xi{}\approx{}0.2-3$. In the following, we adopt $\xi{}=3$.

Dynamical encounters also affect the orbital eccentricity $e$. Following \cite{quinlan1996}, we define the parameter
\begin{equation}\label{eq:kappa}
  \kappa{}\equiv{}\frac{{\rm d}e}{{\rm d}\ln{(1/a)}}.
\end{equation}
Hence, the variation of eccentricity because of dynamical hardening can be expressed as
\begin{equation}\label{eq:kappa2}
  \frac{{\rm d}e}{{\rm d}t}=2\,{}\pi{}\,{}\xi{}\,{}\kappa{}\,{}\frac{G\,{}\rho{}_{\rm c}}{\sigma}\,{}a.
\end{equation}
The values of $\kappa{}$ have been calibrated via $N-$body encounters \citep{quinlan1996,sesana2006}, yielding values $\kappa{}\sim{}0.01-0.1$. Here, we adopt $\kappa{}=0.1$, because large values of $\kappa$ better describe the case in which the ratio between the mass of the BBH and the mass of the intruder is close to one. 

At the same time, a BBH evolves by GW emission, which shrinks the orbit and circularizes it. According to \cite{peters1964}, shrinking and circularization by GW emission are described as:
\begin{eqnarray}\label{eq:peters}
 \frac{{\rm d}a}{{\rm d}t}=-\frac{64}{5}\,{} \frac{G^3 \,{} m_1 \,{} m_2 \,{} (m_1+m_2)}{c^5 \,{} a^3\,{} (1-e^2)^{7/2}} \,{} f_1(e)\nonumber\\
  \frac{{\rm d}e}{{\rm d}t}=-\frac{304}{15}\,{} e \frac{ G^3 \,{} m_1 \,{} m_2 \,{} (m_1+m_2)}{c^5 \,{}a^4 \,{}  (1-e^2)^{5/2}}\,{}f_2(e),
\end{eqnarray}
where
\begin{eqnarray}
f_1(e)=\left(1+\frac{73}{24}\,{}e^2+\frac{37}{96}\,{} e^4\right) \nonumber\\
f_2(e)=\left(1+\frac{121}{304} \,{} e^2\right).
\end{eqnarray}

Putting together eqs.~\ref{eq:hardening}, ~\ref{eq:kappa2} and \ref{eq:peters}, we can describe the evolution of a binary under the combined effect of dynamical hardening and GW emission as \citep{mapelli2018review}:
\begin{eqnarray}\label{eq:mapelli2018}
   \frac{{\rm d}a}{{\rm d}t}=-2\,{}\pi{}\,{}\xi{}\,{}\frac{G\,{}\rho{}_{\rm c}}{\sigma}\,{}a^2-\frac{64}{5}\,{} \frac{G^3 \,{} m_1 \,{} m_2 \,{} (m_1+m_2)}{c^5 \,{} a^3\,{} (1-e^2)^{7/2}}\,{}f_1(e) \nonumber\\
  \frac{{\rm d}e}{{\rm d}t}=2\,{}\pi{}\,{}\xi{}\,{}\kappa{}\,{}\frac{G\,{}\rho{}_{\rm c}}{\sigma}\,{}a-\frac{304}{15}\,{} e \frac{ G^3 \,{} m_1 \,{} m_2 \,{} (m_1+m_2)}{c^5 \,{}a^4 \,{}  (1-e^2)^{5/2}}\,{}f_2(e).\nonumber\\
\end{eqnarray}

{\sc fastcluster} performs the above integration with the Euler scheme and an adaptive time-step. In particular, we reduce (increase) the time-step by a factor of 10 (2) whenever the percentage change of $a$ between two time-steps is $>1$\% ($\leq{}0.1$\%). The code also includes a Runge-Kutta fourth order integrator, if higher accuracy is needed.

For each original binary, we start the integration described by eqs.~\ref{eq:mapelli2018} after the second BH has formed, i.e. after a time $t_{\rm form}$. For each first-generation dynamical binary, we start the integration after the binary has formed by dynamical encounters, i.e. after the time $t_{\rm dyn}$ described by eq.~\ref{eq:tdyn}. 
We integrate  each binary (original or dynamical) for a time equal to $\min{(t_{\rm merg},\,{}t_{\rm Hubble})}$, where $t_{\rm Hubble}\approx{}13.6$ Gyr is the Hubble time. 
For original binaries, $t_{\rm merg}=t_{\rm form}+t_{\rm del}$, where $t_{\rm del}$ is the delay time between the formation of the BBH and its merger. For 1st-generation dynamical binaries, $t_{\rm merg}=t_{\rm dyn}+t_{\rm del}$. We assume that the binary merges when $a\leq{}r_{\rm ISCO}$, where $r_{\rm ISCO}$ is the radius of the innermost stable circular orbit.

At each time-step we compare the current time $t$ with the lifetime of the host star cluster ($t_{\rm SC}$).   If $t\leq{}t_{\rm SC}$, the binary is integrated according to eqs.~\ref{eq:mapelli2018}. If $t>t_{\rm SC}$ the binary no longer hardens and is integrated according to eqs.~\ref{eq:peters}.

At the beginning of the integration of eqs.~\ref{eq:mapelli2018}, we also compare 
the maximum semi-major axis for ejection by three-body encounters \citep{miller2002}:
\begin{equation}
a_{\rm ej}=\frac{2\,{}\xi{}\,{}m_\ast{}^2}{(m_1+m_2)^3}\,{}\frac{G\,{}m_1\,{}m_2}{v_{\rm esc}^2}
\end{equation}
with the maximum semi-major axis for which shrinking by GW emission becomes dominant over shrinking by dynamical hardening (e.g., eq.~23 of \citealt{baibhav2020}):
\begin{equation}
a_{\rm GW}=\left[\frac{32\,{}G^2}{5\,{}\pi{}\,{}\xi{}\,{}c^5}\,{}\frac{\sigma{}\,{}m_1\,{}m_2\,{}(m_1+m_2)}{\rho{}_{\rm c}\,{}(1-e^2)^{7/2}}\,{}f_1(e)\right]^{1/5}.
\end{equation}
 
If $a_{\rm ej}<a_{\rm GW}$, the binary evolves inside the cluster as described by eqs.~\ref{eq:mapelli2018}, until it merges or the star cluster dies by evaporation ($t>t_{\rm SC}$). If the star cluster evaporates before the binary has merged, we keep integrating the orbital evolution of the binary by GW emission, following eqs.~\ref{eq:peters}, up to $\min{(t_{\rm merg},t_{\rm Hubble})}$.

If $a_{\rm ej}\ge{}a_{\rm GW}$, the binary is ejected from the cluster by dynamical recoil before it merges. Inside the main loop of the integration, we switch from eqs.~\ref{eq:mapelli2018}  (dynamical hardening plus GW emission) to eqs.~\ref{eq:peters} (GW emission only) either when the star cluster evaporates  ($t>t_{\rm SC}$) or when the current semi-major axis of the binary $a(t)\leq{}a_{\rm ej}$ (whichever happens first).

This integration is a significant improvement with respect to previous work, which is based on order of magnitude timescales. 
On the other hand, it still contains several approximations. For example, 
we neglect dynamical exchanges, which might change the masses of the binary system.

When a binary merges, we model the mass and spin of the merger remnant using fitting formulas from numerical relativity, as described by \cite{jimenez-forteza2017} (see also \citealt{rezzolla2008,hofmann2016}). 
The final mass is $\approx{}0.9$ the total mass of the two merging BHs, while the final spin magnitude clusters around $\chi_{\rm f}\approx{}0.7-0.9$.

\section{Nth generations}\label{sec:nthgen}

\subsection{Relativistic kicks}\label{sec:kick}

At coalescence, the  merger product receives a kick because of the transfer of linear momentum caused by asymmetries in GW emission  \citep{fitchett1983,lousto2012,maggiore2018}. We model the magnitude of relativistic kicks following \cite{lousto2012}:
\begin{equation}\label{eq:vk}
    v_{\rm kick}=\left(v_{\rm m}^2+v_\perp^2+2\,{}v_{\rm m}\,{}v_\perp\,{}\cos{\phi{}}+v_\parallel^2\right)^{1/2},    
\end{equation}
where
\begin{eqnarray}
  v_{\rm m}=A\,{} \eta{}^2 \,{}\frac{(1-q)}{(1+q)}\,{} (1 + B\,{} \eta{})\nonumber\\
    v_\perp{}=H\,{}\frac{\eta^2}{(1+q)}\,{}\left|\chi_{1\parallel}-q\,{}\chi_{2\parallel}\right|\nonumber\\
    v_\parallel=\frac{16\,{}\eta^2}{(1+q)}\,{}\left[V_{1,1}+V_{\rm A}\,{}S_\parallel+V_{\rm B}\,{}S_\parallel^2+V_{\rm C}\,{}S_\parallel^3\right]\nonumber{}\\
    \,{}\left|\chi_{1\perp{}}-q\,{}\chi_{2\perp{}}\right|\,{}\cos{(\phi_\Delta-\phi)}.\nonumber{}\\    
\end{eqnarray}
In the above equations, $q=m_2/m_1$ with $m_2\le{}m_1$, $\eta{}=q\,{}(1+q)^{-2}$, $A=1.2\times{}10^4$~km~s$^{-1}$, $B=-0.93$, $H=6.9\times{}10^3$~km~s$^{-1}$, $(V_{1,1},\,{}V_{\rm A},\,{}V_{\rm B},\,{}V_{\rm C})=$~(3678, 2481, 1792, 1506) km s$^{-1}$, 
while $\vec{\chi}_1$ and $\vec{\chi}_2$ are the dimensionless spin vectors of the primary and secondary BHs, respectively. Moreover, $\chi_{1\parallel{}}$ ($\chi_{2\parallel}$) is the component of the spin of the primary (secondary) BH parallel to the orbital angular momentum of the binary system, while   $\chi_{1\perp{}}$ ($\chi_{2\perp}$) is the component of the spin of the primary (secondary) BH lying in the orbital plane. $S_\parallel$ is the component parallel to the orbital angular momentum of the vector $\vec{S}=2\,{}(\vec{\chi}_1+q^2\vec{\chi}_2)/(1+q)^2$. Finally, $\phi_\Delta$ represents the angle between the direction of the infall at merger (which we randomly draw in the BBH orbital plane) and the in-plane component of $\vec{\Delta} \equiv{} (m_1 + m_2 )^2\,{} \left(\vec{\chi}_1 - q \,{} \vec{\chi}_2 \right)/(1+q)$, while $\phi$  is the phase of the BBH, randomly drawn between 0 and $2\,{}\pi{}$.

If $v_{\rm kick}<v_{\rm esc}$, the merger product is retained in the star cluster, otherwise it is ejected. If it is ejected, it cannot acquire a new companion by three-body encounters: we do not consider it for the next generations.

\subsection{Relevant timescales and orbital properties}

Even if $v_{\rm kick}<v_{\rm esc}$, the merger remnant is a single BH at birth and it is likely ejected into the outskirts of the star cluster by the relativistic kick. It must sink back to the dense central regions of the star cluster by dynamical friction, before it can acquire a new companion dynamically. Hence, for each merger remnant still bound to the cluster, we first calculate the dynamical friction timescale according to eq.~\ref{eq:tdf}. From now on, there is no difference in the treatment between remnants that originated from the merger of an original BBH or a dynamical BBH.  

After a time $t_{\rm DF}$, the BH is back to the core of the cluster and can acquire a new companion by three-body encounter or by exchange. We then calculate the time needed for the merger remnant to form a new binary with another BH as the minimum between $t_{\rm 3bb}$ (eq.~\ref{eq:t3bb}) and $t_{\rm 12}$ (eq.~\ref{eq:t12}).

The total time to form a second-generation BBH is then
\begin{equation}\label{eq:tdynng}
  t_{\rm dyn,\,{}ng}=t_{\rm merg}+t_{\rm DF}+\min{(t_{\rm 3bb},t_{\rm 12})},
\end{equation}
where $t_{\rm merg}$ is the time to form and merge the previous generation BBH, as described in the previous sections. If $t_{\rm dyn,\,{}ng}<\min{(t_{\rm SC},\,{}t_{\rm Hubble})}$, the second (or nth-) generation BBH can form. We then draw the secondary mass according to eq.~\ref{eq:oleary} and the new eccentricity and semi-major axis as detailed in Section~\ref{sec:dynbin}. The spin of the secondary component is drawn from the distribution of first-generation (nth-generation) BHs if the mass of the secondary is lower (higher) than the maximum mass of first-generation BHs.

We then integrate the evolution of the orbital properties of the nth-generation binary as described in Section~\ref{sec:evolution}, until the minimum between the Hubble time and the merger time $t_{\rm merg}$ of the new binary, which now includes not only the formation time of the binary but also the time-span of the previous generations. If the nth-generation binary merges, we calculate the properties of the merger remnant and the relativistic kick and we re-start the loop from Section~\ref{sec:kick}.

It might be useful to summarize here the relevant timescales used in {\sc fastcluster} and how they work.
\begin{itemize}
\item{}  For dynamical (original) binaries, $t_{\rm form}$ is the time elapsed between the formation of the  progenitor star (binary star) and the formation of the BH (BBH). 
\item{} Eq.~\ref{eq:tdf} defines the dynamical friction timescale $t_{\rm DF}$.
\item{} For  first-generation dynamical binaries and for all nth-generation binaries, $t_{\rm 3bb}$, defined in eq.~\ref{eq:t3bb}, is the timescale for the formation of a BBH by three-body encounters.
\item{}   For  first-generation dynamical binaries and for all nth-generation binaries, $t_{\rm 12}$, defined in eq.~\ref{eq:t12}, is the timescale for the formation of a BBH by dynamical exchange. 
\item{} $t_{\rm dyn}$ is the total time required to form a first-generation dynamical BBH (eq.~\ref{eq:tdyn}).
\item{} $t_{\rm dyn,\,{}ng}$ is the total time required to form a nth-generation BBH (eq.~\ref{eq:tdynng}), including the assembly and evolution of previous generations.
\item{} We define $t_{\rm del}$ as the time elapsed between the formation of the BBH (via either binary evolution or dynamical interactions) and its merger by GW emission. During this time a binary is evolved according to eqs.~\ref{eq:mapelli2018} if it is still inside the star cluster and according to eqs.~\ref{eq:peters} after its dynamical ejection or after the evaporation of the star cluster.
\item{} We define $t_{\rm merg}$ as the total time elapsed from the beginning of the simulation to the merger of a BBH.  For nth-generation BBHs $t_{\rm merg,\,{}ng}$ includes the evolutionary times of the previous generations.
\item{} We indicate the Hubble time as $t_{\rm Hubble}$.
\item{} The lifetime of the star cluster is $t_{\rm SC}$. 
\end{itemize}

\begin{figure*}
  \begin{center}
    \includegraphics[width = 0.85 \textwidth]{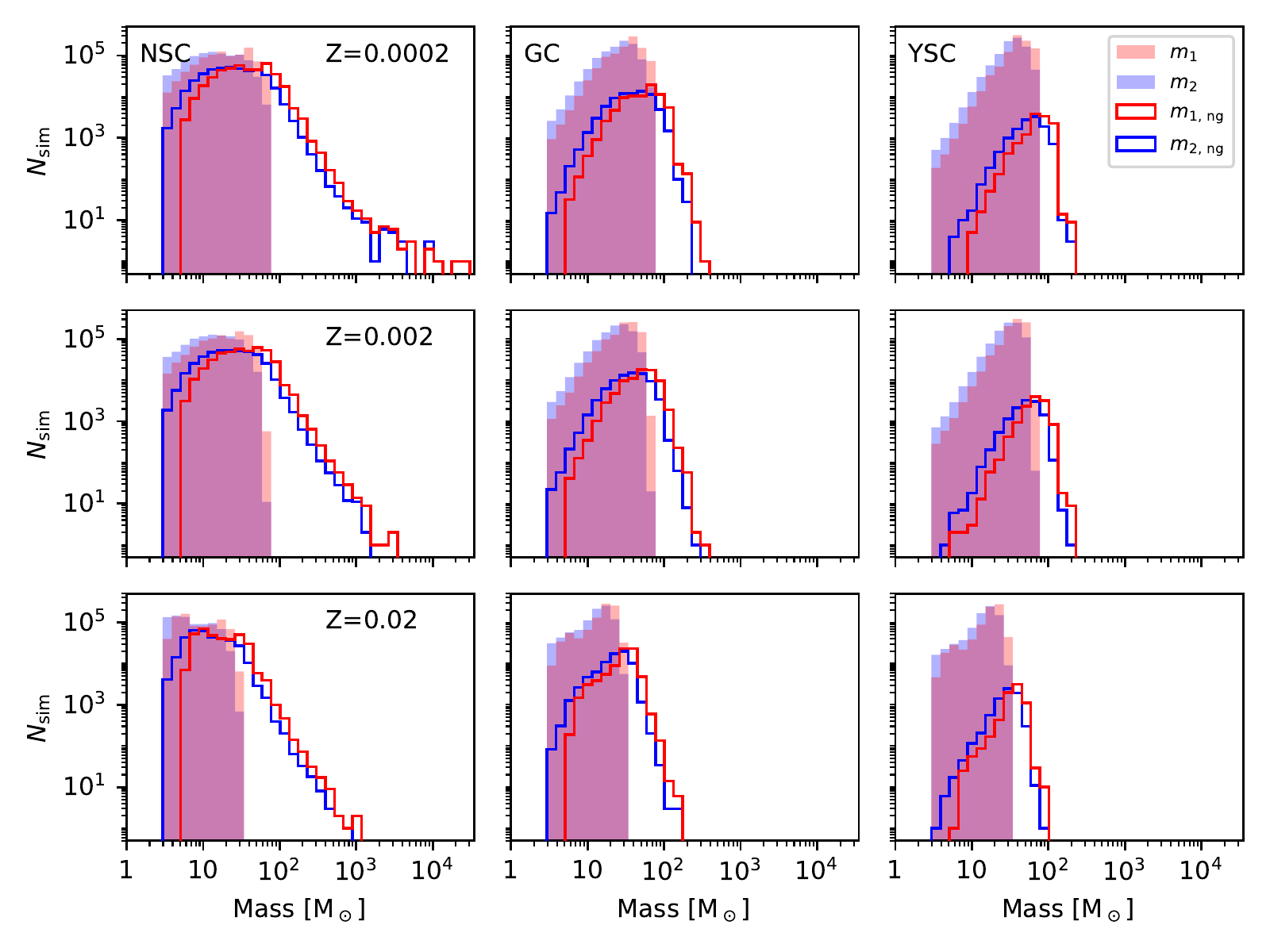}
    \end{center}
  \caption{Masses of the BBH mergers for different stellar metallicities and star clusters. From top to bottom: $Z=0.0002$, 0.002 and 0.02. Left-hand panels: NSCs (from top to bottom: models NSC\_D2, NSC\_D1 and NSC\_D3); middle panels: GCs (models GC\_D2, GC\_D1 and GC\_D3); right-hand panels: YSCs (models YSC\_D2, YSC\_D1 and YSC\_D3). In each panel, the filled red (blue) histogram shows the primary (secondary) mass of first-generation mergers, while the unfilled red (blue) histogram shows the primary (secondary) mass of nth-generation mergers. The $y-$axis shows the number of simulated BBH mergers. 
    \label{fig:compare_met}}
\end{figure*}

\begin{figure*}
  \begin{center}
    \includegraphics[width = 0.85 \textwidth]{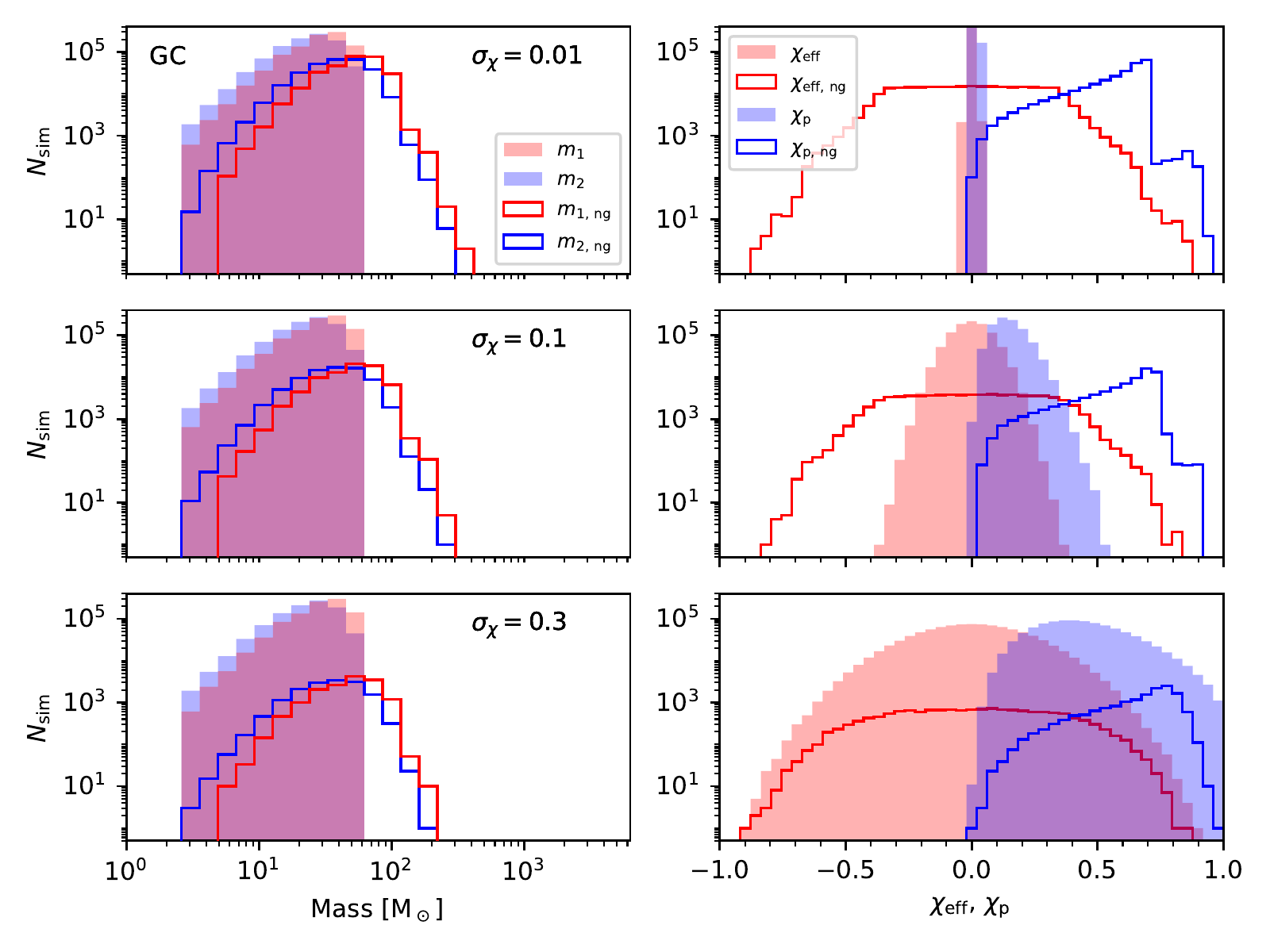}
    \end{center}
  \caption{Masses (left) and spins  (right) of the BBH mergers for different initial spin distributions. The figure shows the case of GCs, but we draw similar conclusions for NSCs and YSCs (not shown for brevity). From top to bottom: first-generation spins drawn from a Maxwellian distribution with one-dimensional root mean square $\sigma{}_\chi=0.01$ (upper panels, model GC\_D4), 0.1 (central panels, fiducial model GC\_D1) and 0.3 (lower panels, model GC\_D5).
    Left-hand panels: the filled red (blue) histogram shows the primary (secondary) mass of first-generation mergers, while the unfilled red (blue) histogram shows the primary (secondary) mass of nth-generation mergers.
    Right-hand panels: the filled red (blue) histogram shows the effective spin (precessing spin) of first-generation mergers, while the unfilled red (blue) histogram shows the effective spin (precessing spin) of nth-generation mergers.
    In all the panels, the $y-$axis shows the number of simulated BBH mergers.
    \label{fig:compare_spin}}
\end{figure*}

\section{Star cluster properties}\label{sec:starcluster}

Most of the aforementioned equations depend on the properties of the host star cluster. Here we assume, for the sake of simplicity, that the star cluster properties do not evolve in time. We will add the evolution of the star cluster in a follow-up study. We consider three different flavours of star clusters: NSCs, GCs and YSCs.

Each star cluster is uniquely defined by its lifetime $t_{\rm SC}$, total mass $M_{\rm tot}$, binary fraction $f_{\rm bin}$ and half-mass density $\rho{}$. We assume $t_{\rm SC}=13.6$, 12 and 1 Gyr for NSCs, GCs \citep{gratton1997,gratton2003,vandenberg2013} and YSCs  \citep{portegieszwart2010}, respectively. Furthermore, we assume $f_{\rm bin}=0.01$, 0.1 and 1 in NSCs \citep{antonini2016}, GCs \citep{jibregman2015} and YSCs \citep{sana2012}, respectively. We draw the total masses  from a log-normal distribution with mean $\langle{}\log_{10}{M_{\rm tot}/{\rm M}_\odot}\rangle{}=6.18,\,{}5.3$ and 4.3 for NSCs, GCs and YSCs, respectively. We assume a fiducial standard deviation $\sigma_{\rm M}=0.4$ for all star cluster flavours. We also consider the cases in which $\sigma_{\rm M}=0.2,$ and 0.6. We draw the  density at the half-mass radius from a log-normal distribution with mean $\langle{}\log_{10}{\rho{}/({\rm M}_\odot\,{}{\rm pc}^{-3})}\rangle{}=5,$ 3.3 and 3.3 for NSCs, GCs and YSCs, respectively. We assume a fiducial standard deviation $\sigma_\rho=0.4$ for all star cluster flavours.  The values of $M_{\rm tot}$ and $\rho{}$ are inferred from the observations reported in \cite{neumayer2020} for NSCs and GCs (see also \citealt{harris1996,georgiev2016}) and from \cite{portegieszwart2010} for YSCs.  For each star cluster, we assume a core density $\rho_{\rm c}=20\,{}\rho$.  We derive the escape velocity from $M_{\rm tot}$ and $\rho$ \citep{georgiev2009a,georgiev2009b,fragione2020} using the following relationship
\begin{equation}\label{eq:vesc}
  v_{\rm esc}=40\,{}{\rm km}\,{}{\rm s}^{-1}\,{}\left(\frac{M_{\rm tot}}{10^5\,{}{\rm M}_\odot}\right)^{1/3}\,{}\left(\frac{\rho}{10^5\,{}{\rm M}_\odot\,{}{\rm pc}^{-3}}\right)^{1/6}.
\end{equation}
 Equation~\ref{eq:vesc} results in a distribution of escape velocities fairly consistent with the observational sample reported in Figure~1 of \cite{antonini2016} for GCs and NSCs.

Here, we do not consider NSCs that host a supermassive BH (SMBH). In such clusters, most of the binaries inside the influence radius of the SMBH are soft and are disrupted over a timescale \citep{binney1987}:
    \begin{equation}
      t_{\rm ev}=\frac{(m_1+m_2)\,{}\sigma{}}{16\,{}\sqrt{\pi}\,{}G\,{}m_\ast{}\,{}\rho{}_{\rm c}\,{}a\,{}\ln{\Lambda}}.
    \end{equation}
We refer to \cite{arcasedda2020b} for a detailed treatment of this case.

\section{Results}\label{sec:results}

\subsection{Description of runs}

We ran 42 different realizations of our models, half of them for original binaries and the other half for dynamical binaries, playing with the most relevant parameters. We consider three different families of star clusters (NSCs, GCs and YSCs), three different metallicities $Z=0.0002,$ 0.002 and 0.02 (roughly corresponding to $0.01,$ 0.1 and 1 Z$_\odot$), three different values of the spin magnitude parameter $\sigma_\chi=0.01,$ 0.1 and 0.3, and three different values of $\sigma_{\rm M}=0.2,$ 0.4 and 0.6. Table~\ref{tab:table1} summarizes the details of each model. Each model consists of $10^6$ first-generation BBHs. In the following section, we describe the main results of these runs.

\subsection{Dynamical binaries in NSCs, GCs and YSCs}

\begin{figure}
  \begin{center}
    \includegraphics[width=0.5 \textwidth]{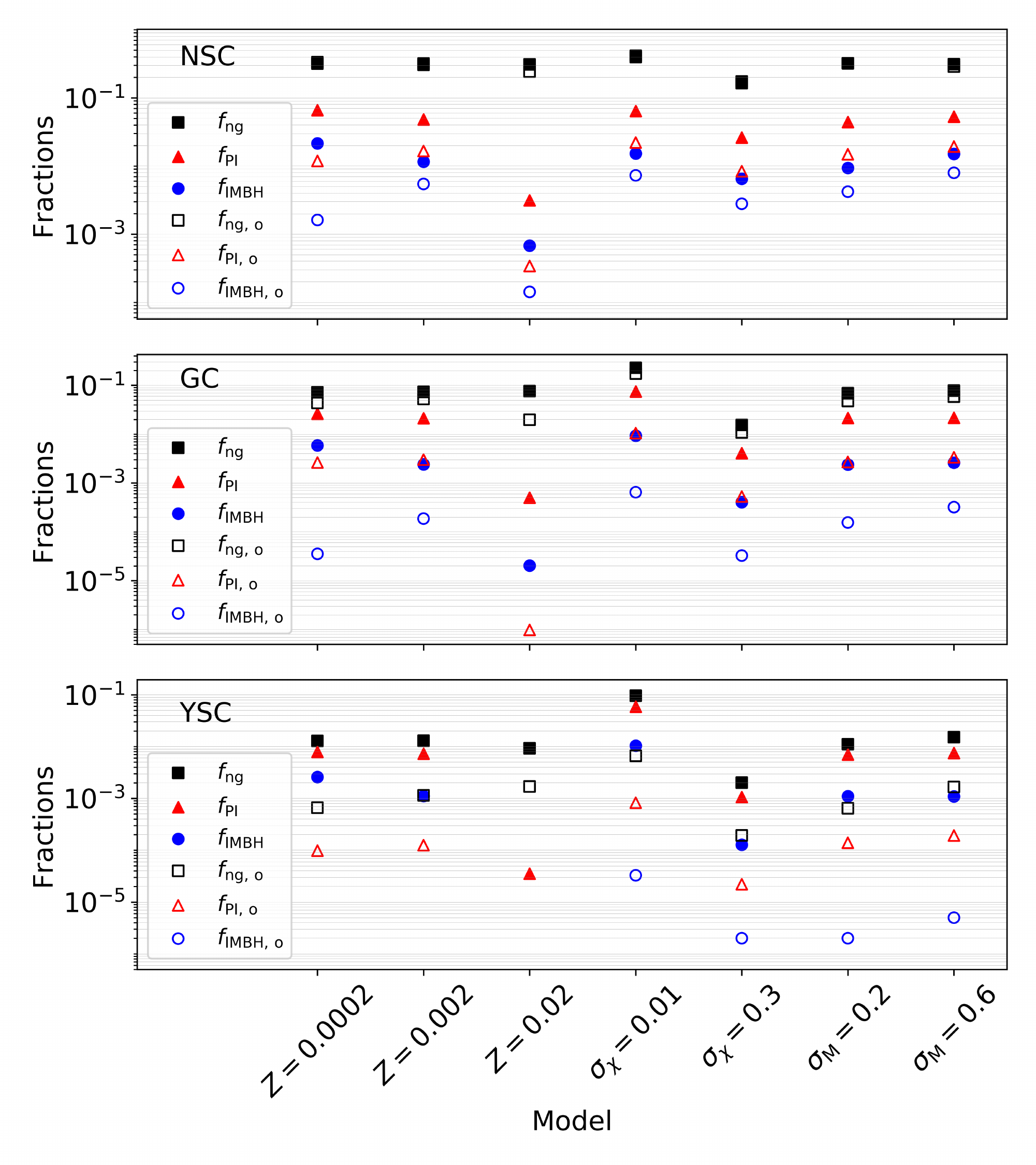}
    \end{center}
  \caption{    Black squares: fraction $f_{\rm ng}$ of nth-generation BBH mergers over the total number of BBH mergers (including both first and nth-generation mergers). Red triangles: fraction $f_{\rm PI}$ of BBH mergers with a primary mass in the PI gap ($65-120$ M$_\odot$) with respect to the total number of BBH mergers. Blue circles: fraction $f_{\rm IMBH}$ of BBH mergers with a primary mass in the intermediate-mass BH regime ($100-10^5$ M$_\odot$) with respect to the total number of BBH mergers. Filled (open) symbols refer to dynamical (original) binaries. The models labeled as $Z=0.0002$ are NSC\_D2, GC\_D2, YSC\_D2, NSC\_O2, GC\_O2 and YSC\_O2, i.e. all the cases with metallicity $Z=0.0002$. Those labeled as $Z=0.002$ are NSC\_D1, GC\_D1, YSC\_D1, NSC\_O1, GC\_O1 and YSC\_O1, i.e. all the fiducial cases with $Z=0.002$, $\sigma_{\chi}=0.1$ and $\sigma_{\rm M}=0.4$. Those labeled as $Z=0.02$ are NSC\_D3, GC\_D3, YSC\_D3, NSC\_O3, GC\_O3 and YSC\_O3, i.e. all the cases with $Z=0.02$. Those labeled as $\sigma_{\chi}=0.01$ ($\sigma_{\chi}=0.3$) are NSC\_D4, GC\_D4, YSC\_D4, NSC\_O4, GC\_O4 and YSC\_O4 (NSC\_D5, GC\_D5, YSC\_D5, NSC\_O5, GC\_O5 and YSC\_O5), i.e. all the cases with $\sigma_{\chi}=0.01$ ($\sigma_{\chi}=0.3$). Those labeled as $\sigma_{\rm M}=0.2$ ($\sigma_{\rm M}=0.6$) are NSC\_D6, GC\_D6, YSC\_D6, NSC\_O6, GC\_O6 and YSC\_O6 (NSC\_D7, GC\_D7, YSC\_D7, NSC\_O7, GC\_O7 and YSC\_O7), i.e. all the cases with $\sigma_{\rm M}=0.2$ ($\sigma_{\rm M}=0.6$). 
    \label{fig:frac}}
\end{figure}

\begin{figure}
  \begin{center}
    \includegraphics[width=0.5 \textwidth]{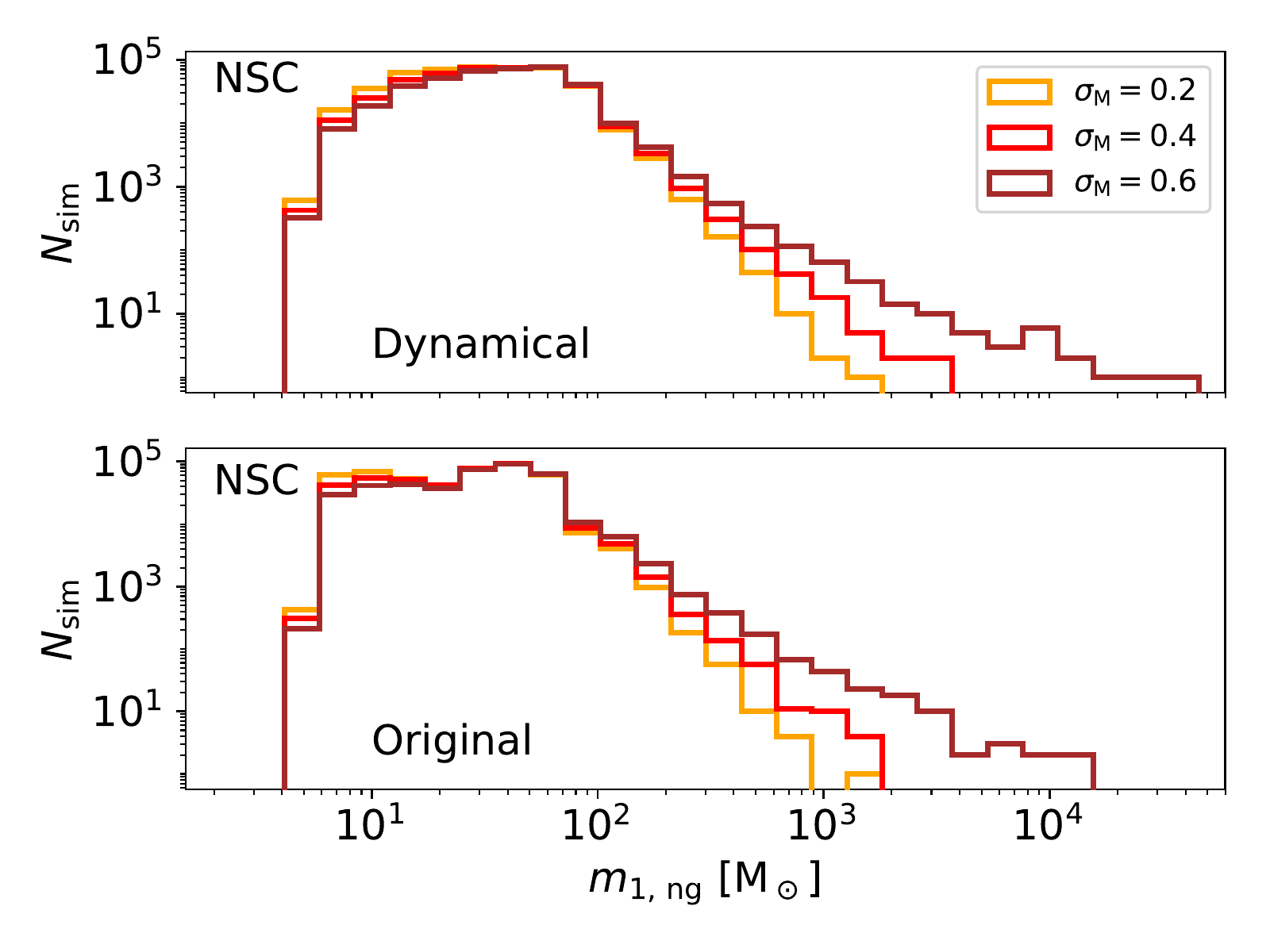}
    \end{center}
  \caption{
    Primary mass distribution for nth-generation BBHs in NSCs. Upper panel: dynamical BBHs; lower panel: original BBHs. Orange line: models with $\sigma_{\rm M}=0.2$ (NSC\_D6 and NSC\_O6 in the upper and lower panel, respectively); light red line: models with $\sigma_{\rm M}=0.4$ (NSC\_D1 and NSC\_O1 in the upper and lower panel, respectively); dark red line: models with $\sigma_{\rm M}=0.6$ (NSC\_D7 and NSC\_O7 in the upper and lower panel, respectively).
    \label{fig:vesc}}
\end{figure}

Figures~\ref{fig:NSCdyn}, \ref{fig:GCdyn} and \ref{fig:YSCdyn} show the main properties of dynamical BBHs formed in NSCs, GCs and YSCs, respectively, according to our fiducial models (NSC\_D1, GC\_D1 and YSC\_D1). Only 
BBHs that merge within a Hubble time  are shown, both first- and nth-generation. This figure does not distinguish between BBHs that merge inside or outside their parent star cluster.

The mass function of first-generation BBHs is similar in NSCs, GCs and YSCs, because it comes from the same catalogues, 
but for one crucial difference: the assumption that only BHs receiving a natal kick lower than the escape velocity have a chance to pair up dynamically (Section~\ref{sec:dynbin}) prevents the formation of low-mass BBHs especially in the clusters with the lowest escape velocity. Hence, NSCs witness the formation of more low-mass BBHs than both GCs and YSCs. For this reason, the mass function of first-generation and nth-generation dynamical BBHs includes a larger fraction of low-mass systems ($m_1<20$ M$_\odot$) in NSCs than in both GCs and YSCs (see also Table~\ref{tab:table2}). 

The mass function of nth-generation BBHs peaks at values $\sim{}30-100$ M$_\odot$ in all considered star clusters, indicating that most nth-generation BBHs are just second generation. The main difference between NSCs, GCs and YSCs is the maximum mass of nth-generation BBHs. NSCs, because of their high escape velocity, allow a larger number of generations to form, up to primary masses in excess of $\sim{}10^3$ M$_\odot$ (Table~\ref{tab:table2}). The main limitation to build even more massive BHs in NSCs is represented by the long timescales: after $\approx{}10$ generations at most, the simulation reaches the Hubble time. In contrast, the maximum masses in both GCs and YSCs are a few times $10^2$ M$_\odot$. Another crucial difference between NSCs and either GCs or YSCs is the fraction of nth- to first-generation mergers (Table~\ref{tab:table2}).

The distribution of spins looks similar in NSCs, GCs and YSCs, by construction, because we assume the same spin models.  The main feature is the double horned distribution of the precessing spin $\chi_{\rm p}$: most first-generation mergers have precessing spin squeezed toward low values ($\sim{}0.1-0.2$), while nth-generation mergers tend to have high values of $\chi_{\rm p}\sim{}0.7$. This creates a sort of spin gap between $\sim{}0.3$ and $\sim{}0.6$, as already discussed in \cite{baibhav2020} and \cite{fishbach2020}. The importance of the secondary peak at large $\chi_{\rm p}$ with respect to the primary peak at $\chi_{\rm p}\sim{}0.1-0.2$ depends on the fraction of nth- to first-generation mergers. 

The eccentricity distributions when the binary enters the LIGO--Virgo band (i.e., when the frequency of GW emission is $f_{\rm GW}=10$ Hz) are similar for NSCs, GCs and YSCs, because the dynamical evolution is comparable for the three samples. The fraction of dynamical BBHs with eccentricity $e>0.1$ ($e>0.9$) at $f_{\rm GW}=10$ Hz is $\sim{}2\times{}10^{-3}$ ($\sim{}10^{-4}$) for all types of star clusters, when accounting for both first and nth-generation mergers (Table~\ref{tab:table2}).

Finally, the dynamical formation timescales of first- and nth-generation dynamical BBHs ($t_{\rm dyn}$ and $t_{\rm dyn,\,{}ng}$) are clearly squeezed to lower values for YSCs ($t_{\rm dyn}\sim{}3-20$ Myr) than for both GCs ($t_{\rm dyn}\sim{}20-200$ Myr) and NSCs (peaking at $t_{\rm dyn}\sim{}20-200$ Myr, but with a considerable fraction of systems with $t_{\rm dyn}>1$ Gyr). Indeed, dynamical friction timescales, three-body timescales and exchange timescales are usually shorter in YSCs, which have lower velocity dispersion than both GCs and YSCs, but relatively high central density and large binary fractions  \citep{portegieszwart2010,neumayer2020}.

\subsection{Original binaries in NSCs, GCs and YSCs}

Figures~\ref{fig:NSCoriginal}, \ref{fig:GCoriginal} and \ref{fig:YSCoriginal} show the same distributions as Figure~\ref{fig:NSCdyn}, but for original binaries in NSCs, GCs and YSCs, respectively (NSC\_O1, GC\_O1 and YSC\_O1). The main differences with respect to dynamical binaries concern the distribution of first-generation masses and eccentricities and the relevant timescales.

The masses of first-generation original BBHs are skewed to lower values than those of first-generation dynamical BBHs, because supernova kicks are not as efficient in ejecting binary systems from the parent star cluster and because the maximum mass for the components of a BBH merger is only $\sim{}45$ M$_\odot$ in original binaries. This implies that the bulk of nth-generation mergers descending from original binaries is also shifted to lower values with respect to dynamical binaries: $\sim{}20-70$ M$_\odot$ instead of $\sim{}30-100$ M$_\odot$, at $Z=0.002$. As a consequence, the maximum masses of nth-generation mergers are also shifted to lower values. Even in the case of original binaries, the fraction of nth-generation to first-generation mergers is considerably higher in NSCs than GCs and YSCs.

Even if three-body encounters tend to increase the eccentricity of binary systems, the eccentricities of first-generation original mergers are squeezed to lower values, with almost no systems with eccentricity $>0.1$ in the LIGO--Virgo band. Nth-generation mergers are the only ones to populate this region. 
Table~\ref{tab:table2} shows that this is true not only for the three fiducial models of original binaries (NSC\_O1, GC\_O1 and YSC\_O1), but also for all the other models with one exception: the high metallicity case. In the models NSC\_O3, GC\_O3 and YSC\_O3 (original binaries with progenitor metallicity $Z=0.02$) the percentage of mergers with eccentricity $e>0.1$ is $\approx{}1-2$\%. This is not an effect of dynamical encounters, but a consequence of supernova kicks. Most of the highly eccentric systems at solar metallicity are first-generation mergers, which form with high eccentricity because of the supernova kick they receive at birth, right after a common-envelope phase \citep{giacobbo2018}.

While spin distributions are the same in original and dynamical binaries by construction, the relevant timescales are remarkably different because the formation time of an original BBH is always between $t_{\rm form}\approx{}3$ and $\approx{}20$ Myr, corresponding to the formation time of the second BH. In contrast, the time $t_{\rm dyn}$ for the dynamical assembly of a first-generation BBH varies wildly depending on the properties of both the primary BH and the host star cluster: it could be as short as $\sim{}3$ Myr or as long as $\sim{}t_{\rm Hubble}$.

\subsection{Impact of metallicity on masses}

The metallicity of the progenitors leaves a strong imprint on the mass spectrum of first-generation and, consequently, nth-generation mergers, for both original and dynamical binaries. Figure~\ref{fig:compare_met} shows the primary and secondary masses of dynamical BBHs. The peak of the nth-generation BHs tends to cluster around the maximum mass of first-generation mergers. At solar metallicity this maximum mass is $\approx{}30$ M$_\odot$, while it rises up to  $\approx{}70$ M$_\odot$ for dynamical binaries at low metallicity.

Table~\ref{tab:table2} shows the median value of primary BHs' mass in first and nth-generation mergers for all the considered models, together with the maximum mass. In NSCs, the median value of primary BH mass in first-generation dynamical binaries is $\approx{}20$ M$_\odot$ at low metallicity ($Z=0.002,$ 0.0002) and $\approx{}8$ M$_\odot$ at solar metallicity, while the median value in nth-generation mergers is $\approx{}35$ ($\approx{}15$)~M$_\odot$ at low (solar) metallicity.

In GCs and YSCs, the median values are considerably larger than in NSCs, for both first-generation and nth-generation dynamical BHs, because (as we mentioned before) the lightest single BHs are easily ejected by supernova kicks and cannot pair up dynamically. Hence, in GCs the median mass of first-generation primary BHs is $\approx{}30$ ($\approx{}17$) M$_\odot$ at low (solar) metallicity, while the median mass of nth-generation primary BHs is $\approx{}50$ ($\approx{}30$) M$_\odot$ at low (solar) metallicity. In YSCs, the median mass of first-generation primary BHs is $\approx{}40$ ($\approx{}19$) M$_\odot$ at low (solar) metallicity, while the median mass of nth-generation primary BHs is $\approx{}70$ ($\approx{}40$) M$_\odot$ at low (solar) metallicity.

We can draw similar conclusions for original binaries, just shifted to lower mass values. For the sake of brevity, we do not show an additional figure for original binaries, but Table~\ref{tab:table2} clearly shows that the median masses of primary BHs are lower in original binaries. For example, in NSCs, the median mass of primary BHs in first-generation original binaries is $9-10$ M$_\odot$ at low metallicity and $6$ M$_\odot$ at solar metallicity. The same quantity for primary BHs of nth-generation binaries is $24-32$ M$_\odot$ at low metallicity and only 9 M$_\odot$ at solar metallicity.

\subsection{Impact of spin distribution}

The actual distribution of BH spin magnitudes is highly uncertain from both theory \citep[e.g.,][]{belczynski2020,bavera2020} and observations \citep[e.g.,][]{miller2015,abbottO3popandrate}. For this reason, we explore a wide range of values for $\sigma_{\chi}$. Figure~\ref{fig:compare_spin} shows the impact of different spin distributions on the mass of BBHs for dynamical binaries in GCs. We do not show NSCs, YSCs and original binaries because the effect is similar to the one shown in Figure~\ref{fig:compare_spin}. Different spin magnitudes do not significantly impact the shape of the mass function, the maximum mass and the position of the peak, but they have a strong effect on the number of nth-generation mergers.

This effect is particularly important for YSCs and GCs, which have a lower escape velocity than NSCs. As shown by Figure~\ref{fig:frac} and by Table~\ref{tab:table2}, the fraction of nth-generation mergers $f_{\rm ng}$ over all BBH mergers is only $\approx{}0.02$  of the total mergers for $\sigma_{\chi}=0.3$ and for dynamical binaries in GCs, while it rises to $\approx{}0.075$ for $\sigma_{\chi}=0.01$. This happens because larger spins are associated with larger relativistic kicks than smaller spin magnitudes, for a given mass ratio and inclination of the spins.

Figure~\ref{fig:compare_spin} also shows that the double horned shape of the distribution of $\chi_{\rm p}$ disappears for the large spin case ($\sigma_{\chi}=0.3$), while it is a strong feature of both the low-spin and intermediate-spin cases.

\subsection{Impact of star cluster mass}

Escape velocity is possibly the key quantity to  drive the number and properties of nth-generation mergers. Most of the differences between YSCs, GCs and NSCs spring from the different $v_{\rm esc}$, with a larger escape velocity leading to more nth-generation mergers, and to a higher maximum mass. The parameter that mainly affects the escape velocity is star cluster mass ($v_{\rm esc}\propto{}M_{\rm tot}^{1/3}$, eq.~\ref{eq:vesc}).

Figure~\ref{fig:vesc} shows the distribution of primary masses of nth-generation BBH mergers in NSCs when changing the standard deviation of the total star cluster mass distribution. The good news is that the changes do not affect the position of the peak, which corresponds to masses $m_{\rm 1,\,{}ng}=30-80$ M$_\odot$ for both dynamical and original BBHs. On the other hand, a larger standard deviation leads to higher possible maximum masses, associated to the most massive clusters. In the case NSC\_D7 ($\sigma_{\rm M}=0.6$), the maximum BH mass is $\approx{}4.4\times{}10^4$ M$_\odot$, while it is only $\approx{}1500$ M$_\odot$ in NSC\_D6 ($\sigma_{\rm M}=0.2$). Such massive nth-generation BBHs are extremely rare anyway ($\sim{}1$ every $10^6$ BBH mergers).

Table~\ref{tab:table2} shows not only the maximum primary BH mass but also the total number of generations per each model. The effect of varying $\sigma_{\rm M}$ on the number of generations $N_{\rm g}$ is apparent, especially for NSCs: $N_{\rm g}=17,$ 10 and 8 in NSC\_D7 ($\sigma_{\rm M}=0.6$), NSC\_D1 ($\sigma_{\rm M}=0.4$) and NSC\_D6 ($\sigma_{\rm M}=0.2$), respectively. 

\subsection{Efficiency of hierarchical mergers}

There is a remarkable difference between NSCs and the other types of star clusters, if we look at the efficiency of hierarchical mergers. Here, for efficiency we mean both i) the fraction of nth-generation mergers with respect to all BBH mergers ($f_{\rm ng}$) and ii) the maximum number of generations ($N_{\rm g}$) achieved in a given model. Both $f_{\rm ng}$ and $N_{\rm g}$ are listed in Table~\ref{tab:table2}. Figure~\ref{fig:frac} shows the behaviour of $f_{\rm ng}$ across different models. 

In NSCs, $f_{\rm ng}$ is almost always $\gtrsim{}0.3$, with the exception of the models NSC\_O3 (original binaries with solar metallicity),   NSC\_D5 and NSC\_O5 (with $\sigma_{\chi}=0.3$). In particular, $f_{\rm ng}$ is a factor of $\sim{}2$ lower for the two high-spin cases. The maximum number of generations is always $N_{\rm g}\geq{}8$. Overall, NSCs are very efficient in producing hierarchical mergers because of their high escape velocity, with just a mild dependence on the other considered parameters, especially the spin magnitudes. 

In GCs, $f_{\rm ng}$ changes more wildly, from $\sim{}0.01$ to $\sim{}0.2$. Most GC models with dynamical (original) binaries have $f_{\rm ng}\approx{}0.07-0.08$ ($\approx{}0.04-0.06$). Models with low and high spins are both outliers: the two low-spin models GC\_D4 and GC\_O4 have $f_{\rm ng}\approx{}0.23$ and $\approx{}0.17$, respectively, while the two high-spin cases GC\_D5 and GC\_O5 have $f_{\rm ng}\approx{}0.02$ and $\approx{}0.01$. The maximum number of generations ranges from 4 to 6 in all the considered models. Overall, the lower escape velocity of GCs makes the spin magnitude parameter a crucial one to decide the efficiency of hierarchical mergers in GCs.

Finally, YSCs have the lowest efficiency of hierarchical mergers, as expected because of both the lower escape velocity and the shorter lifetime (1 Gyr). In this case, we also see a conspicuous difference between dynamical and original binaries. Most YSC models with dynamical (original) binaries have $f_{\rm ng} \sim{}0.01$  ($f_{\rm ng}\sim{}10^{-3}$), with a difference of about one order of magnitude between dynamical and original binaries. This is a consequence of the much lower first-generation masses of original binaries with respect to dynamical binaries (Table~\ref{tab:table2}), which makes it even more difficult for the nth-generation systems to merge within a Hubble time, in addition to the short lifetime and low escape velocity of YSCs. Similar to GCs, the spin distribution plays a major role to further suppress or enhance the fraction of nth-generation mergers in YSCs: $f_{\rm ng}\approx{}0.096$ and $0.002$ according to models YSC\_D4 ($\sigma_{\chi}=0.01$) and YSC\_D5 ($\sigma_{\chi}=0.3$). The maximum number of generations is always between 3 and 4.

To summarize, the efficiency of hierarchical mergers is always larger in NSCs than either GCs or YSCs, mostly as a result of the escape velocity. Large (small) spins tend to suppress (enhance) the efficiency of hierarchical mergers, but their impact is much larger for both GCs and YSCs than for NSCs. Original binaries are less efficient in  leading to the hierarchical growth with respect to the more massive dynamical binaries. Overall, NSCs are less sensitive to the main parameters (including spin magnitudes) with respect to both GCs and YSCs.

Even if we can reach up to 17 generations in the most lucky case (NSC\_D7), the fraction of BBH mergers in each generation with respect to the total number of BBH mergers decreases very fast. The lower panel of Figure~\ref{fig:ngen_mmax} shows that the fraction of mergers in the last generation 
is lower than $10^{-4}$. The second generation is always at least one order of magnitude more populated than the next one.

\subsection{BHs in the PI gap and intermediate-mass BHs}

\begin{figure}
  \begin{center}
    \includegraphics[width=0.5 \textwidth]{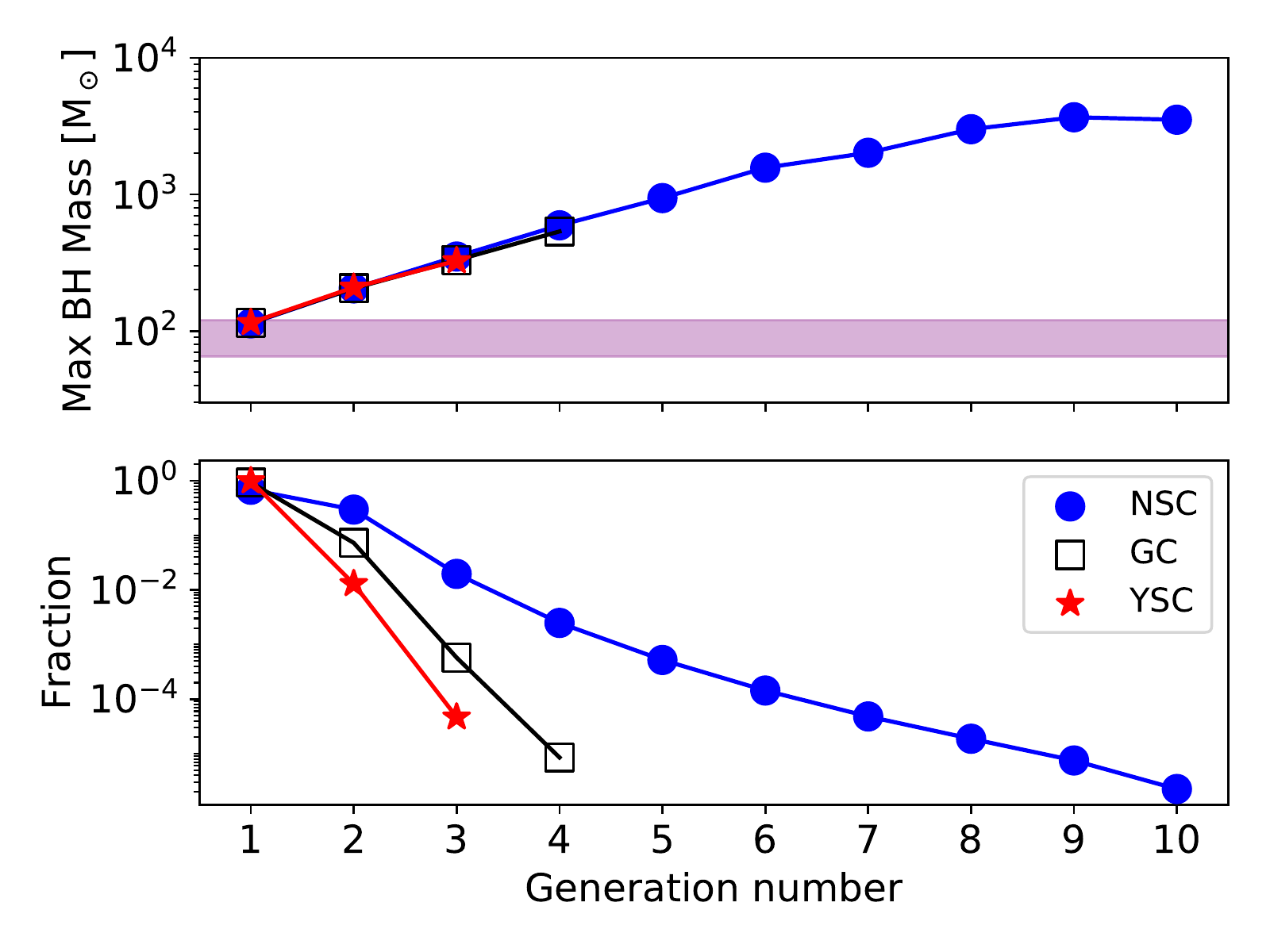}
    \end{center}
  \caption{Upper panel: maximum merger remnant mass in each generation as a function of the generation number (where 1 means first generation). Lower panel: Fraction of mergers belonging to a given generation with respect to all BBH mergers in the considered model as a function of the generation number. Red stars: YSCs from the fiducial model for dynamical BBHs (YSC\_D1). Open black squares: GCs from the fiducial model for dynamical BBHs (GC\_D1). Filled blue circles: NSCs from the fiducial model for dynamical BBHs (NSC\_D1). 
    \label{fig:ngen_mmax}}
\end{figure}

\begin{figure}
  \begin{center}
    \includegraphics[width=0.5 \textwidth]{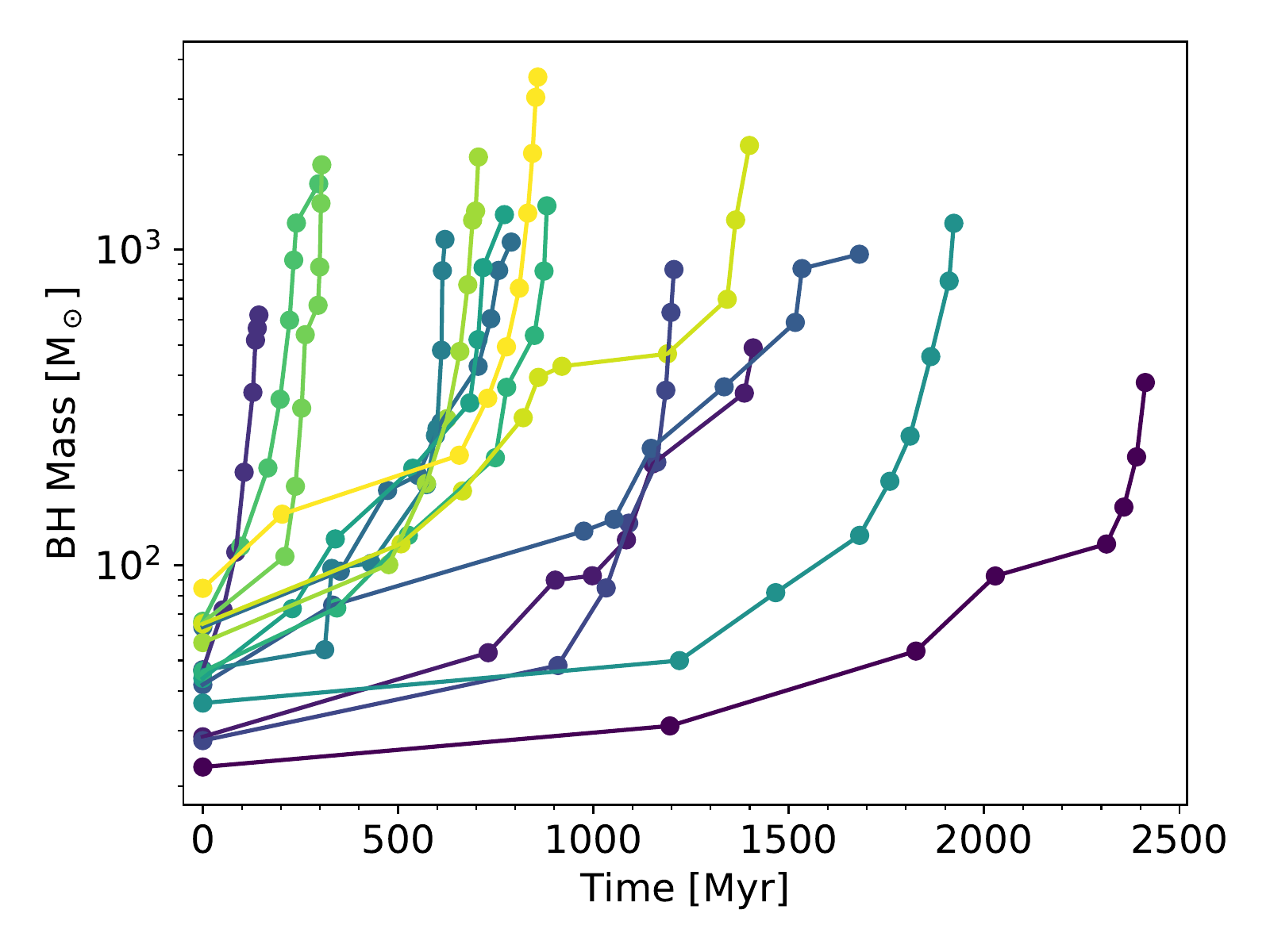}
    \end{center}
  \caption{Time evolution of the BH remnant mass for 15 randomly selected hierarchical mergers reaching at least the 8th generation in model NSC\_D1. Each circle marks a merger. We set to zero the time of the first-generation merger. 
    \label{fig:BH_growth}}
\end{figure}

Hierarchical mergers are one of the most likely scenarios to form BHs in the PI mass gap and even intermediate-mass BHs (IMBHs). Indeed, we have probably already witnessed the formation of an IMBH via the merger of two smaller BHs (GW190521, \citealt{abbottGW190521,abbottGW190521astro}). Our synthetic models show that the population of BHs in the mass gap and IMBHs is extremely sensitive to the metallicity of the progenitor stars and to the properties of the host star cluster (mass and density).

Figure~\ref{fig:frac} shows the fraction of BBH mergers that have at least the primary mass in the mass gap ($f_{\rm PI}$) with respect to all BBH mergers of first- and nth-generation. The values of $f_{\rm PI}$ are also listed in Table~\ref{tab:table2}. We calculated $f_{\rm PI}$ by assuming the mass gap to be between $65$ and $120$ M$_\odot$, given the uncertainties on its boundaries \citep[e.g.,][]{farmer2019,farmer2020,mapelli2020,farrell2020,tanikawa2020,renzo2020,vanson2020,costa2020}. 

While the fraction of nth-generation mergers $f_{\rm ng}$ is almost insensitive to stellar metallicity, $f_{\rm PI}$ varies wildly depending on this parameter. For example, if we consider dynamical binaries in NSCs, $f_{\rm PI}$ decreases from $\sim{}0.05-0.07$ to $\sim{}0.003$ if we go from the metal-poor ($Z=0.0002,$ 0.002) to the metal-rich ($Z=0.02$) cases. The reason for this feature is that the mass of first-generation BHs is crucial to determine how many second-generation BHs form with mass in the PI gap, and the mass of first-generation BHs strongly depends on progenitors' metallicity in our models.

For the same reason, $f_{\rm PI}$ is considerably lower for original binaries than for dynamical binaries. This happens because the mass of first-generation original mergers is lower than the mass of first-generation dynamical binaries. For example, $f_{\rm PI}\approx{}0.048$ and 0.017 in NSC\_D1 and NSC\_O1, respectively.

GCs and YSCs are generally associated with lower values of $f_{\rm PI}$ with respect to NSCs. For example, in the fiducial dynamical case, $f_{\rm PI}\approx{}0.048,$ 0.021 and 0.007 in NSCs, GCs and YSCs, respectively. There is one exception, though: if the spin magnitudes are very low, $f_{\rm PI}$ is comparable in all types of star clusters: $f_{\rm PI}\approx{}0.064,$ 0.075 and 0.059 in NSC\_D4, GC\_D4 and YSC\_D4, respectively. This happens for the same reason as for $f_{\rm ng}$: when the spins are very low (as our $\sigma_{\chi}=0.01$ case), the retention fraction of second-generation BHs dramatically rises in lower mass clusters and becomes comparable to NSCs.

We can draw similar statements for the fraction ($f_{\rm IMBH}$) of merging binaries that include at least one IMBH over all BBH mergers. We consider IMBHs all the simulated BHs with mass $\geq{}100$ M$_\odot$, according to an historical definition \citep{millercolbert2004}. Figure~\ref{fig:frac} and Table~\ref{tab:table2} show the values of $f_{\rm IMBH}$ for all the considered cases. $f_{\rm IMBH}$ can be as high as $\approx{}0.02$ in the case of dynamical binaries in NSCs and as low as 0 in the case of original binaries in YSCs. It is higher (lower) in metal-poor (metal-rich) star clusters and when we consider dynamical (original) binaries. The value of $f_{\rm IMBH}$ is particularly high in GCs and YSCs when the low-spin case is considered. For example, $f_{\rm IMBH}$ rises from $\approx{}0.001$ in the model YSC\_D1 to $\approx{}0.01$ in YSC\_D4.

The upper panel of Figure~\ref{fig:ngen_mmax} shows the maximum mass of the merger remnant in each generation for the fiducial case (NSC\_D1, GC\_D1 and YSC\_D1). The maximum remnant mass is already in the IMBH regime after the first generation. There is not much difference between NSCs, GCs and YSCs if we consider the same generation number. 
NSCs end building up larger BHs only because they can witness a larger number of generations with respect to both YSCs and GCs. The maximum remnant mass in the most extreme generations ($n\sim{}3$ for YSCs, $\sim{}4$ for GCs and $\sim{}10$ for NSCs) is close to $\sim{}10^{-3}$ $M_{\rm tot}$ (where $M_{\rm tot}$ is the total cluster mass), analogous to some previous numerical results \citep[e.g.,][]{portegieszwart2002,giersz2015,dicarlo2019,kroupa2020} and reminiscent of the observational relationship between the mass of a NSC and that of its central BH in the considered mass range \citep[e.g.,][]{graham2009,neumayer2020}. Finally, Figure~\ref{fig:BH_growth} shows the time evolution of the remnant mass in model NSC\_D1 for 15 randomly selected systems. The duration of the hierarchical growth ranges from $\approx{}100$ Myr to several Gyr in NSCs.

\begin{figure}
  \begin{center}
    \includegraphics[width=0.5 \textwidth]{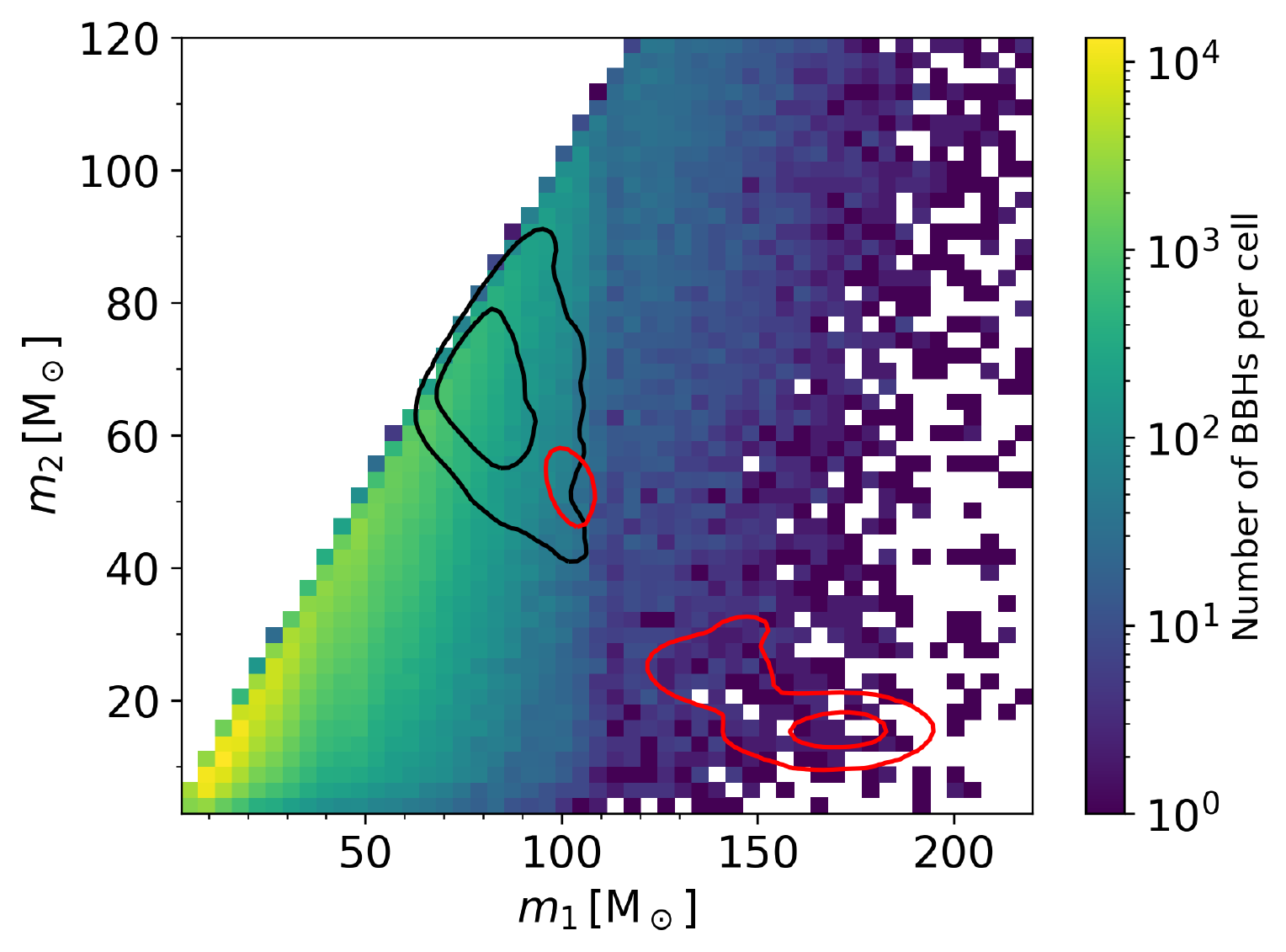}
        \includegraphics[width=0.5 \textwidth]{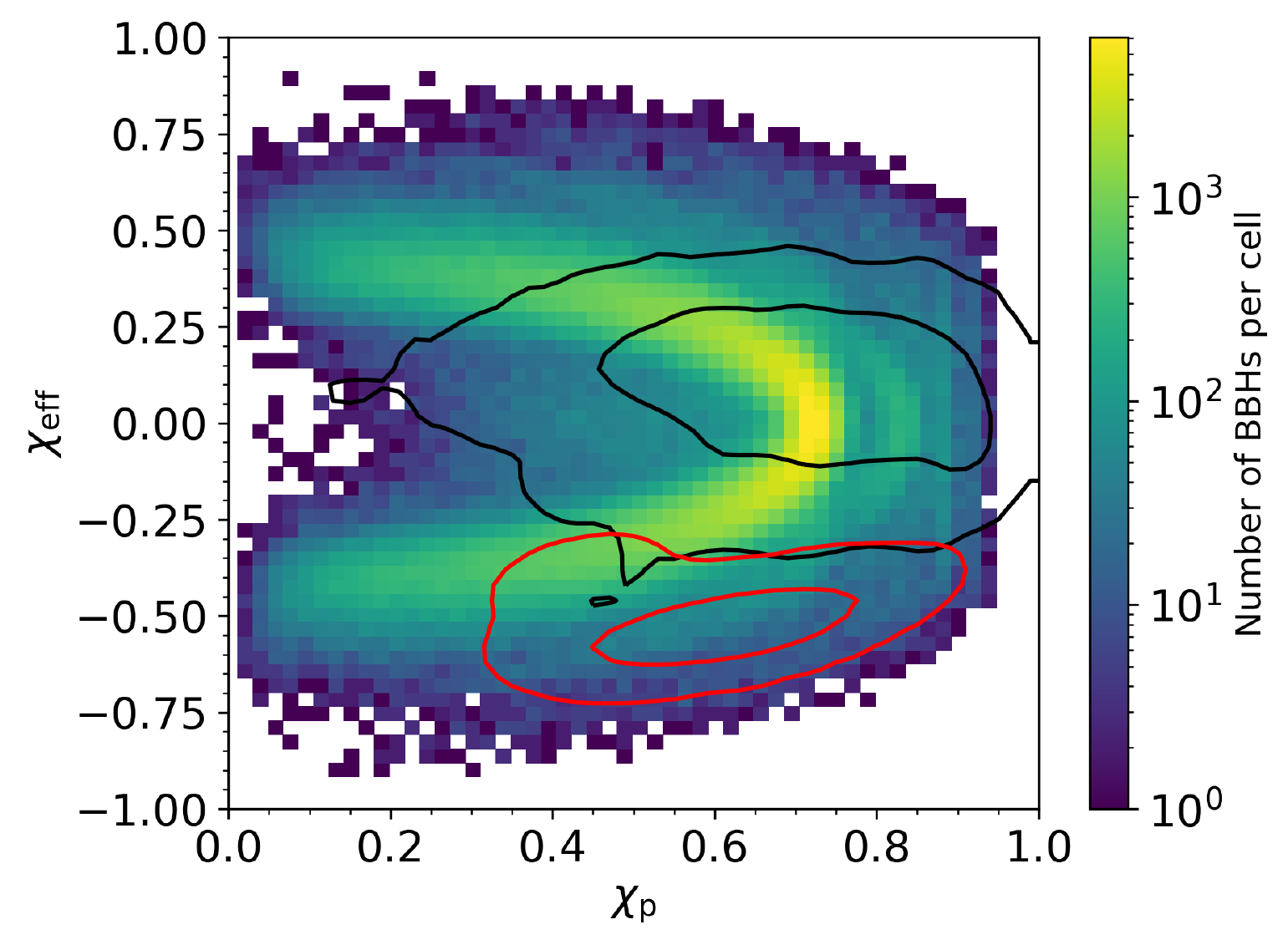}
    \end{center}
  \caption{Upper panel: Mass of secondary versus primary component of nth-generation BBH mergers in model NSC\_D1. Lower panel: effective versus  precessing spin of nth-generation BBH mergers in NSC\_D1. In both plots, the contours show the 90 and 50\% credible intervals for GW190521 according to {\protect\cite{abbottGW190521}}, in black, and {\protect\cite{nitz2021}}, in red, respectively.}
    \label{fig:GW190521}
\end{figure}

\subsection{GW190521: inside the gap or intermediate-mass  ratio inspiral?}

We explore the possibility that a system like GW190521 forms from our hierarchical models. Previous studies have already investigated the formation of GW190521 via hierarchical mergers \cite[e.g.,][]{abbottGW190521astro,fragione2020quad,fragione2020b,rizzuto2020,fishbach2020, gayathri2020,deluca2020,liulai2021, Rice2020, romero-shaw2020,palmese2020,
 safarzadeh2020,samsing2020}. Also, \cite{kimball2020a} find that GW190521 is favored to contain two second-generation BHs with odds $>700$. Our results confirm  that this scenario is able to match both the masses and the spins of GW190521 (Table~\ref{tab:table2}). Recently, \cite{nitz2021} reanalyzed the data of the LIGO--Virgo collaboration (LVC) with a new waveform allowing for more extreme mass ratios. They find that GW190521 is consistent with an intermediate-mass ratio inspiral with primary mass $m_1=168^{+15}_{-61}$~M$_\odot$ and secondary mass $m_2=16^{+33}_{-3}$~M$_\odot$ (within 90\% credible interval, according to a uniform in mass-ratio prior). In this case, the primary mass is not inside the PI mass gap. We have calculated how many of our systems have both masses and spin parameters inside the 90\% credible interval inferred by \cite{nitz2021} and  by \cite{abbottGW190521}, respectively. As shown by Table~\ref{tab:table2} and Figure~\ref{fig:GW190521}, our results indicate that the properties of GW190521-like systems inferred from the analysis of \cite{nitz2021} are more difficult to match by the hierarchical merger scenario than the properties inferred by \cite{abbottGW190521}. For example, in the model NSC\_D1, the fraction of GW190521-like systems over  all possible BBH mergers is $f_{\rm GW190521}\approx{}0.02$  and $\approx{}4\times{}10^{-4}$ if we use the mass and spin parameters from \cite{abbottGW190521} and from \cite{nitz2021}, respectively. Hence, the new estimates from \cite{nitz2021} pose GW190521 on the "safe side" with respect to PI theory, but might be even more difficult to explain with an astrophysical model.  Our assumption for the mass distribution of secondary BHs (eq.~\ref{eq:oleary}), motivated by previous star cluster simulations, strongly influences this result.
 
 Finally, \cite{gayathri2020} and \cite{romero-shaw2020} propose that GW190521 is consistent with an eccentric BBH merger. Among all the systems that match the mass and spin of GW190521 in our models, the fraction of those that have eccentricity $e>0.5$ in the LIGO--Virgo band is 
 $\sim{}10^{-3}$ in our fiducial dynamical models. However, our current formalism does not include GW captures \citep{zevin2019}. Hence, the fraction of systems with extreme eccentricity we estimated should be regarded as a lower limit.

\section{Discussion}\label{sec:discussion}

We studied the evolution of hierarchical mergers born from both dynamical and original binaries. The latter have been overlooked in previous work. In our treatment, we did not attempt to quantify the relative weight of original versus dynamical binaries. Are dynamical or original binaries more representative of the actual population of BBH mergers in star clusters? A reasonable guess is that YSCs should be  dominated by original binaries, because they have a binary fraction close to one \citep[e.g.,][]{sana2012}. In contrast,  NSCs are probably dominated by dynamical binaries, because most original binaries are too soft and get ionized, while GCs should stay in between. In a follow-up study, we will try to constrain the relative fraction of original versus dynamical binaries in different star clusters. Overall, the main differences between a star cluster dominated by original binaries and one dominated by dynamical binaries are i) the mass distribution of first-generation BBHs (more squeezed towards lower values for original binaries), ii) the global distribution of eccentricities in the LIGO--Virgo band (original binaries inherit lower eccentricities from binary star evolution), iii) the efficiency of hierarchical mergers and formation of BBHs in the PI gap or IMBHs.

The formalism presented here might lead to an overestimation of the differences between original and dynamical binaries. In fact, even original binaries can undergo dynamical exchanges before the two original members of the binary merge. Dynamical exchanges are currently not implemented in our model and tend to lead to the formation of more and more massive binary systems, which are more energetically stable \citep{hills1980}. Hence, exchanges might reduce the differences between the mass function of original and dynamical binaries.

Moreover, we have not included stellar and binary evolution. This leads us to neglect star--star collisions and runaway mergers, which are triggered by dynamical encounters  and can be an additional formation channel of BHs in the PI mass gap and IMBHs, especially in metal-poor YSCs \citep[][]{dicarlo2019,dicarlo2020a,dicarlo2020b,kremer2020,renzo2020b}. Given their short dynamical friction and core collapse timescales, YSCs are particularly efficient in forming massive BHs via runaway collisions \citep[e.g.,][]{portegieszwart2004,giersz2015,mapelli2016}. Hence, including stellar collisions might bridge the gap between YSCs and more massive clusters, leading to comparable values of $f_{\rm PI}$ and $f_{\rm IMBH}$.

 We have assumed that the properties of a star cluster do not evolve during its life.  On the one hand, star clusters lose mass by stellar evolution and dynamical ejection and expand by two-body relaxation. This leads to lower star cluster mass and density, possibly quenching the formation of hierarchical mergers \citep[e.g.,][]{antonini2020a, antonini2020b}. On the other hand, by assuming no evolution with time, we do not account for core collapse episodes and gravothermal oscillations, which lead to a dramatic temporary increase of the central density, possibly boosting BBH formation and hierarchical mergers \citep[e.g.,][]{breen2013}. NSCs might even acquire mass during their life by fresh star formation \citep[e.g.,][]{mapelli2012} and by accreting GCs \citep[e.g.,][]{capuzzo2008,antonini2012}. These processes might lead to a higher efficiency of hierarchical mergers in NSCs. The overall effect of including star cluster evolution in our model is thus quite difficult to predict and might be significantly different for YSCs, GCs and NSCs. We will add a formalism for star cluster evolution in a follow-up study.

Furthermore, we assumed that each BBH, after formation, remains inside the cluster core until it merges or it is ejected from the cluster. This might overestimate the effect of dynamical hardening because the binary may wander around the cluster as an effect of three-body encounters and Brownian motion \citep[e.g.,][]{arcasedda2020b}.

Several previous studies have investigated hierarchical mergers either with simulations or semi-analytic models. We briefly compare our main results against some relevant previous work. The possibility of growing large IMBHs (up to $\sim{}4\times{}10^4$ M$_\odot$ in our simulations) when particularly massive ($>10^7$ M$_\odot$) NSCs are considered is in good agreement with previous semi-analytic models \citep[e.g.,][]{antonini2019,fragione2020,mapelli2020b}.   Similar to \cite{baibhav2020}, we recover the prediction of a spin gap between first-generation and nth-generation mergers, when we consider relatively low values of the spin parameter ($\sigma_{\chi}=0.01$, 0.1) for first-generation BHs. 

The strong impact of the initial spin magnitude distribution on both $f_{\rm ng}$ and $f_{\rm PI}$ confirms a previous result by \cite{rodriguez2019}, based on dynamical simulations. They find that $\sim{}10$\% ($\sim{}1$\%) of the BBH mergers in their simulated GCs have mass in the PI gap when a constant value of $\chi{}=0$ ($\chi=0.5$) is assumed. For comparison, in our models GC\_D4 ($\sigma_{\chi}=0.01$) and GC\_D5 ($\sigma_{\chi}=0.3$), we obtain $f_{\rm PI}\approx{}0.075$ and $0.004$, respectively. These numbers and the results of \cite{rodriguez2019} are in fair agreement, if we take into account the differences among the two methods (e.g., a constant value versus a Maxwellian distribution for the spin magnitudes). With respect to both \cite{rodriguez2019} and \cite{zevin2019} we find significantly less systems with extreme eccentricity in the LVC band. This happens because we do not include GW captures in our model. In this sense, our eccentricity distribution should be regarded as a lower limit.

\cite{kimball2020a} study the properties of BBH mergers in GWTC-2  with a phenomenological population model, optimized for GCs (see also \citealt{kimball2020} and \citealt{doctor2020}). They conclude that the  rate of mergers between a second-generation and a first-generation BH is about one order of magnitude lower than the one of first-generation BBHs. It is almost impossible to make a one-to-one comparison between our results and these models, because of the intrinsic differences of the methodology, but the order of magnitude is the same as the values of $f_{\rm ng}$ reported in Table~\ref{tab:table2} for most NSC and GC models.

In summary, our new tool compares quite well with the results of previous semi-analytic models and dynamical simulations. It is particularly flexible, because we can start from different binary catalogues, based on either population synthesis or phenomenological models, and is considerably fast: we can integrate the hierarchical merger of $\approx{}10^6$ initial binaries per single core per day. This will allow a future exploration of an even larger parameter space than the one considered here.

\section{Summary}\label{sec:summary}

We investigated the hierarchical merger scenario with a new fast synthetic tool: {\sc fastcluster} evolves a population of BBHs in star clusters, taking into account both hardening by three-body encounters and GW decay. The first-generation BBHs we considered in this work have two possible formation channels: original and dynamical BBHs. The former descend from hard binary stars, while the latter assembly via dynamical friction, three-body encounters and dynamical exchanges. We evolve the first-generation and nth-generation binaries in three different environments: YSCs, GCs and NSCs. We explore different progenitor metallicities ($Z=0.0002,$ 0.002 and 0.02), spin magnitude distributions ($\sigma_{\chi}=0.01$, 0.1 and 0.3) and star cluster mass distributions ($\sigma_{\rm M}=0.2$, 0.4 and 0.6).

First-generation dynamical binaries have larger average masses than original binaries, because only BHs with mass up to $\sim{}45$ M$_\odot$ can form in hard original binaries, as a consequence of non-conservative mass transfer and common envelope, while dynamical binaries tend to host more massive first-generation BHs (up to $\sim{}65$ M$_\odot$), which  harden faster.

The median values of the primary BH mass in dynamical BBH mergers are larger in both YSCs  and GCs  than in NSCs (see Table~\ref{tab:table2}). This seemingly odd result is a consequence of supernova kicks. BHs ejected by supernova kicks cannot pair up dynamically. In YSCs and GCs, the lighter BHs are ejected 
by the supernova kick and do not participate to the dynamical assembly. In contrast, NSCs, which have a much higher escape velocity, retain a larger fraction of light BHs.

The bulk of the population of nth-generation BHs has mass comparable with the maximum mass of first-generation BH mergers. Progenitor's metallicity has a strong impact on the typical mass of first- and nth-generation BBHs, both in original and dynamical binaries. For example, in our fiducial case for dynamical binaries in NSCs (NSC\_D1), the median mass of the primary BH in first-generation (nth-generation) mergers is $\approx{}20$ M$_\odot$ ($\approx{}35$ M$_\odot$) in metal-poor clusters, while it drops to $\approx{}8$ M$_\odot$ ($\approx{}15$ M$_\odot$) at solar metallicity.

The maximum possible BH mass is much larger in NSCs than in both YSCs and GCs because of the higher escape velocity, which allows to build up a larger number of generations. In our fiducial cases, we form BHs with mass up to a few thousand M$_\odot$ in NSCs and up to a few hundred M$_\odot$ in both GCs and YSCs. NSCs host up to 10 generations in the fiducial case (and up to 17 in the most optimistic case), while GCs and YSCs typically witness up to $4-5$ and $3-4$ generations, respectively.

Original binaries tend to have lower eccentricity than dynamical binaries,  
because of the impact of binary evolution processes. Nth-generation mergers are all the result of dynamical assembly and tend to compensate this initial difference between first-generation dynamical and original binaries. Overall, large eccentricities in the LIGO--Virgo band are extremely rare. For example, the fraction of mergers with eccentricity $e>0.1$ ($e>0.5$) in the LIGO--Virgo band is $\approx{}0.002$ ($10^{-4}$) for the dynamical binaries and $6\times{}10^{-4}$ ($10^{-4}$) for the original binaries in our fiducial case for NSCs, when we consider all the generations together.

Spin magnitudes affect mostly the number of nth-generation mergers with respect to first-generation. This effect is particularly strong in YSCs and GCs. NSCs are less sensitive to spins, because of their higher escape velocity. The most relevant feature in the spin distribution of the global population (including both first-generation and nth-generation mergers) is the double horned distribution of the precessing spin $\chi_{\rm p}$. This shape appears in the models where we assume that first-generation spin magnitudes are small. For example, in our fiducial cases, first-generation mergers peak at $\chi_{\rm p}\sim{}0.1-0.2$, while nth-generation mergers peak at $\chi_{\rm p}\sim{}0.7$.

We calculate the fraction $f_{\rm ng}$ of nth-generation mergers over all possible  BBH mergers. $f_{\rm ng}$ is of the order of $\approx{}0.3$ ($\approx{}0.3$), $\approx{}0.07$ ($\approx{}0.05$) and $\approx{}0.01$ ($\approx{}0.001$) for our fiducial model of dynamical (original) binaries in NSCs, GCs and YSCs, respectively. Hence, the nth-generation fraction is maximum for NSCs, where it is also rather insensitive to the considered parameters. In GCs and YSCs, this fraction is highly sensitive to the spin magnitude; for example, in GCs $f_{\rm ng}$ is up to $\sim{}0.23$ if the spin parameter $\sigma_{\chi}=0.01$ and down to $\sim{}0.02$  if $\sigma_{\chi}=0.3$. This happens because larger spins lead to stronger relativistic kicks, able to eject the merger remnant. NSCs are less sensitive to this difference ($f_{\rm ng}$ changes only by a factor of two between the low spin and the high spin case) because of their high escape velocity.

Hierarchical mergers efficiently trigger the formation of BBHs with primary mass in the PI mass gap or in the IMBH regime. In our fiducial cases for dynamical binaries (with $Z=0.002$), the fraction of mergers with primary mass in the PI gap is $f_{\rm PI}\approx{}0.048$, 0.021 and 0.007 for NSCs, GCs and YSCs, respectively. In the same models, the fraction of mergers with primary mass in the IMBH regime is $f_{\rm IMBH}\approx{}0.012$, 0.002 and 0.001 for NSCs, GCs and YSCs, respectively.

Unlike the global fraction of nth-generation mergers, the fraction of BBHs in the PI mass gap and in the IMBH regime is dramatically affected by metallicity (Table~\ref{tab:table2}). For example,  $f_{\rm PI}$ drops by a factor of $\approx{}40$ in GCs, if we go from $Z=0.002$ to $Z=0.02$. Spins are also important, lower spins enhancing the fraction of mergers in the mass gap and in the IMBH regime with respect to larger spins.

Finally, we investigated the possibility that GW190521 is the result of a hierarchical merger. A recent re-analysis of the LVC data \citep{nitz2021} indicates that this event might be explained by the merger between an IMBH ($\sim{}168$ M$_\odot$) and a $\sim{}16$ M$_\odot$ companion. We estimate the fraction of simulated mergers that match the mass and spin of GW190521 within the 90\% credible interval if we assume the values from \cite{abbottGW190521} and from \cite{nitz2021}, respectively. In all our models, it is much easier to produce a system matching the parameters derived by \cite{abbottGW190521} with respect to \cite{nitz2021}. For example, in the model NSC\_D1 the fraction of BBH mergers inside the 90\% credible interval of GW190521 are $\approx{}0.019$ and $4\times{}10^{-4}$ if we take masses and spins from \cite{abbottGW190521} and \cite{nitz2021}, respectively.  The interpretation proposed by \cite{nitz2021} does not require a violation of the PI mass gap, but might be even more difficult to explain with current astrophysical models.

\section*{Acknowledgements}

 We thank the anonymous referee for their insightful comments, which helped improve this work. We also thank Isobel Romero-Shaw for useful discussions. 
MM, AB, YB, UNDC, NG, GI and FS acknowledge financial support from the European Research Council for the ERC Consolidator grant DEMOBLACK, under contract no. 770017. MCA and MM acknowledge financial support from the Austrian National Science Foundation through FWF stand-alone grant P31154-N27. NG is supported by Leverhulme Trust Grant No. RPG-2019-350 and Royal Society Grant No. RGS-R2-202004. MAS acknowledges financial support from the Alexander von Humboldt Foundation for the research program ``The evolution of black holes from stellar to galactic scales'', the Volkswagen Foundation Trilateral Partnership through project No. I/97778, and the Deutsche Forschungsgemeinschaft (DFG) -- Project-ID 138713538 -- SFB 881.

\section*{Data availability}

The data underlying this article will be shared on reasonable request to the corresponding authors.

\begin{table*}
	\begin{center}
	\caption{Main results of the models presented in this paper.\label{tab:table2}}
	\begin{tabular}{cccccccccccc}
          		\toprule
  Model & $f_{\rm ng}$ & $f_{\rm PI}$ & $f_{\rm IMBH}$  & Med $m_1$  & Med $m_{\rm 1,ng}$  & Max $m_{\rm 1,ng}$  & $N_{\rm g}$ & $f(e>0.1)$ & $f(e>0.9)$ & $f_{\rm GW190521}$ & $f_{\rm GW190521}$\\
   &  &  &   & $\left[{\rm M}_\odot\right]$ & $\left[{\rm M}_\odot\right]$ & $\left[{\rm M}_\odot\right]$ &  &  & & LVC & Nitz \& Capano\\
\midrule
NSC\_D1 & 0.318 & 0.048 & 0.012 & 19 & 34 & 3040  & 10 & 0.002 &  $9\times{}10^{-5}$ & 0.019 & $4\times{}10^{-4}$\\
 NSC\_D2 & 0.319 & 0.066 & 0.021 & 20 & 35 & 28390 & 15 &0.002 &  $8\times{}10^{-5}$ & 0.014 &  $2\times{}10^{-4}$ \\
 NSC\_D3 & 0.308 & 0.003 & $7\times{}10^{-4}$ & 8   & 15 & 1092 & 10 & 0.001 &  $6\times{}10^{-5}$ &  $6\times{}10^{-4}$ &  $8\times{}10^{-6}$\\
 NSC\_D4 & 0.414 & 0.064 & 0.015 & 19 & 34 & 2742  & 10 & 0.002 &  $8\times{}10^{-5}$ & 0.019 &  $3\times{}10^{-4}$\\
 NSC\_D5 & 0.174 & 0.026 & 0.006 & 19 & 35 & 3312  & 11 & 0.002 & $8\times{}10^{-5}$ & 0.007 & $10^{-4}$\\
 NSC\_D6 & 0.321 & 0.044 & 0.009 & 18 & 30 & 1486  & 8 & 0.002 & $9\times{}10^{-5}$ & 0.013 &  $2\times{}10^{-4}$\\
 NSC\_D7 & 0.312 & 0.053 & 0.015 & 21 & 38 & 43587 & 17 & 0.002 & $7\times{}10^{-5}$ & 0.015 &  $2\times{}10^{-4}$\vspace{0.2cm}\\
 GC\_D1  & 0.073 & 0.021 & 0.002 & 31 & 51 & 302   & 4 & 0.002 & $9\times{}10^{-5}$ & 0.008 &  $10^{-4}$\\
 GC\_D2  & 0.072 & 0.026 & 0.006 & 35 & 58 & 341   & 4 & 0.002 & $9\times{}10^{-5}$ & 0.006 &  $2\times{}10^{-5}$\\
 GC\_D3  & 0.077 & $5\times{}10^{-4}$ & $2\times{}10^{-5}$ & 17 & 31 & 164 & 4 & 0.002 & $8\times{}10^{-5}$ & $10^{-4}$ & 0 \\
 GC\_D4  & 0.228 & 0.075 & 0.009 & 31 & 53 & 343   & 5 & 0.002 & $9\times{}10^{-5}$ & 0.024 & $10^{-4}$\\
 GC\_D5  & 0.015 & 0.004 & $4\times{}10^{-4}$ & 31 & 49  & 197 & 4 & 0.002 & $10^{-4}$ & 0.001 & $3\times{}10^{-6}$\\
 GC\_D6  & 0.069 & 0.021 & 0.002 & 31 & 52 & 278   & 4 & 0.002 & $9\times{}10^{-5}$ & 0.007 & $2\times{}10^{-5}$\\
 GC\_D7  & 0.078 & 0.022 & 0.003 & 32 & 50 & 426   & 5 & 0.002 & $9\times{}10^{-5}$ & 0.007 & $3\times{}10^{-5}$ \vspace{0.2cm}\\
 YSC\_D1 & 0.013 & 0.007 & 0.001 & 38 & 69 & 189   & 3 & 0.002 & $10^{-4}$ & 0.003 & $4\times{}10^{-5}$\\
 YSC\_D2 & 0.013 & 0.008 & 0.003 & 42 & 76  & 226  & 3 & 0.002 & $10^{-4}$ & 0.003 & $3\times{}10^{-6}$\\
 YSC\_D3 & 0.009 & $3\times{}10^{-5}$ & 0.0 & 19 & 37 & 86 & 3 & 0.002 & $9\times{}10^{-5}$ & $10^{-5}$ & 0\\
 YSC\_D4 & 0.096 & 0.059 & 0.010 & 38 & 71 & 293   & 4 & 0.002 & $10^{-4}$ & 0.021 & $5\times{}10^{-5}$ \\
YSC\_D5  & 0.002 & 0.001 & $10^{-4}$ & 38 & 66 & 147 & 3 & 0.002 & $10^{-4}$& 0.003 & 0\\
 YSC\_D6 & 0.011 & 0.007 & 0.001 & 38 & 71 & 238   & 4 & 0.002 & $10^{-4}$ & 0.002 & $5\times{}10^{-6}$\\
 YSC\_D7 & 0.015 & 0.008 & 0.001 & 38 & 65 & 319   & 4 & 0.002 & $10^{-4}$ & 0.003 & $10^{-5}$\vspace{0.2cm}\\
NSC\_O1  & 0.305 & 0.017 & 0.005 & 9 &  29 & 1783  & 10 & $6\times{}10^{-4}$ & $10^{-4}$ & 0.003 & $3\times{}10^{-5}$\\
 NSC\_O2 & 0.332 & 0.012 & 0.002 & 10 & 24 & 2903  & 10 &  $6\times{}10^{-4}$ & $4\times{}10^{-5}$ & 0.002 & $9\times{}10^{-5}$\\
 NSC\_O3 & 0.243 & $3\times{}10^{-4}$ & $10^{-4}$    & 6 & 9 & 751 & 10 & 0.014 & 0.014 & $6\times{}10^{-5}$ & $3\times{}10^{-6}$\\
 NSC\_O4 & 0.396 & 0.022 & 0.007 & 9 & 30 & 4784   & 12 &  $8\times{}10^{-4}$ & $10^{-4}$ & 0.003 & $10^{-4}$\\
 NSC\_O5 & 0.164 & 0.008 & 0.003 & 9 & 26 & 1546   & 9 &  $4\times{}10^{-4}$ & $10^{-4}$ & 0.001 & $6\times{}10^{-5}$ \\
 NSC\_O6 & 0.320 & 0.015 & 0.004 & 9 & 26 & 1285   & 9 &  $7\times{}10^{-4}$ & $10^{-4}$ & 0.002 & $9\times{}10^{-5}$\\
 NSC\_O7 & 0.287 & 0.019 & 0.008 & 9 & 32 & 11796  & 13 &  $6\times{}10^{-4}$ & $10^{-4}$ & 0.003 & $10^{-4}$\vspace{0.2cm}\\
 GC\_O1  & 0.053 & 0.003 & $2\times{}10^{-4}$ & 9 & 40 & 225  & 4 &  $10^{-4}$ & $3\times{}10^{-5}$ & $9\times{}10^{-4}$& $10^{-6}$\\
 GC\_O2  & 0.044 & 0.003 & $4\times{}10^{-5}$ & 10 & 31 & 203 & 4 &  $10^{-4}$ & $3\times{}10^{-5}$& $2\times{}10^{-4}$& $3\times{}10^{-6}$\\
 GC\_O3  & 0.020 & $10^{-6}$ & 0.0     & 6  & 8  & 68  & 4 & 0.018 & 0.018& 0 & 0 \\
 GC\_O4  & 0.174 & 0.010  & $6\times{}10^{-4}$ & 9 & 41 & 236 & 5 &  $4\times{}10^{-4}$ & $3\times{}10^{-5}$ & $6\times{}10^{-4}$& $2\times{}10^{-5}$\\
 GC\_O5  & 0.011 & $5\times{}10^{-4}$ & $3\times{}10^{-5}$ & 9  & 39 & 149 & 4 &  $3\times{}10^{-5}$ & $8\times{}10^{-6}$& $2\times{}10^{-5}$& $2\times{}10^{-6}$\\
 GC\_O6  & 0.048 & 0.003 & $2\times{}10^{-4}$ & 9 & 41 & 213 & 4 &  $10^{-4}$ & $10^{-5}$& $10^{-4}$& $10^{-6}$\\
 GC\_O7  & 0.058 & 0.003 & $3\times{}10^{-4}$ & 9 & 39 & 308 & 6 &   $10^{-4}$ & $10^{-5}$\vspace{0.2cm}& $2\times{}10^{-4}$& $8\times{}10^{-6}$\\
 YSC\_O1 & 0.001 & $10^{-4}$ & 0.0 & 9 & 53 & 99 & 3 &  $2\times{}10^{-5}$ & $2\times{}10^{-5}$ & $4\times{}10^{-5}$& 0\\
 YSC\_O2 & $7\times{}10^{-4}$ & $10^{-4}$ & 0.0 & 10 &  41 & 85 & 3 &  $3\times{}10^{-5}$ & $3\times{}10^{-5}$ & $4\times{}10^{-6}$& 0\\
 YSC\_O3 & 0.002 & 0.0     & 0.0 & 6  &   8 & 21 & 3 & 0.022  & 0.022 & 0 & 0\\
 YSC\_O4 & 0.007 & $10^{-4}$ & $3\times{}10^{-5}$ & 9 & 54 & 124 & 3 &  $7\times{}10^{-5}$ & $5\times{}10^{-5}$ & $3\times{}10^{-5}$& $10^{-6}$\\
 YSC\_O5 & $2\times{}10^{-4}$ & $2\times{}10^{-5}$ & $2\times{}10^{-6}$ & 9 & 53 & 113 & 3 & $4\times{}10^{-5}$ & $4\times{}10^{-5}$& 0 & 0\\
 YSC\_O6 & $6\times{}10^{-4}$ & $10^{-4}$ & $2\times{}10^{-6}$ & 9 & 60 & 116 & 3 &  $3\times{}10^{-5}$ & $3\times{}10^{-5}$& $5\times{}10^{-6}$ & 0\\
 YSC\_O7 & 0.002 & $2\times{}10^{-4}$ & $5\times{}10^{-6}$ & 9 & 51 & 128 & 3 &  $3\times{}10^{-5}$ & $3\times{}10^{-5}$ & $5\times{}10^{-6}$ & 0\vspace{0.1cm}\\
		\bottomrule
	\end{tabular}
	\end{center}
	\footnotesize{Column 1: Name of the model; Column 2 ($f_{\rm ng}$): fraction of nth-generation mergers with respect to all BBH mergers; Column 3 ($f_{\rm PI}$): fraction of BBH mergers with at least one component in the PI mass gap with respect to all BBH mergers; Column 4 ($f_{\rm IMBH}$): fraction of BBH mergers with at least one  IMBH component with respect to all BBH mergers; Column 5 (Med $m_1$): median value of the mass of the primary component of a first-generation BBH merger; Column 6 (Med $m_{\rm 1,ng}$): median value of the mass of the primary component of a nth-generation BBH merger; Column 7 (Max $m_{\rm 1,ng}$): maximum mass of the primary component of a nth-generation BBH merger; Column 8 ($N_{\rm g}$): maximum number of generations; Column 9 ($f(e>0.1)$): fraction of BBHs with orbital eccentricity $>0.1$ in the LIGO--Virgo band (i.e., when $f_{\rm GW}=10$ Hz); Column 10 ($f(e>0.5)$): fraction of BBHs with orbital eccentricity $>0.5$ in the LIGO--Virgo band; Columns 11 and 12 ($f_{\rm GW190521}$): fraction of BBHs that match the mass and spin of GW190521, inside the 90\% credible interval, when considering \cite{abbottGW190521} and \cite{nitz2021}, respectively.}
\end{table*}

\bibliographystyle{mnras}
\bibliography{bibliography}

\bsp	
\label{lastpage}
\end{document}